\documentclass[reprint,
 amsmath,amssymb,
 aps,
]{revtex4-2}

\usepackage{xcolor}
\usepackage{graphicx}
\usepackage{dcolumn}
\usepackage{bm}
\usepackage{hyperref}

\usepackage[normalem]{ulem} 
\newcommand{\cor}{\textcolor{black}} 

\begin{document}

\preprint{}

\title{Dynamical structures in phase-separating non-reciprocal polar active mixtures}

\author{Kim L. Kreienkamp}
\email{k.kreienkamp@tu-berlin.de}
\author{Sabine H. L. Klapp}
\email{sabine.klapp@tu-berlin.de}
\affiliation{%
 Institut f\"ur Theoretische Physik, Hardenbergstra\ss e 36, Technische Universit\"at Berlin, D-10623 Berlin, Germany 
}%


\begin{abstract}
Non-reciprocal systems exhibit diverse dynamical phases whose character depends on the type and degree of non-reciprocity. In this study, we theoretically investigate dynamical structures in a mixture of non-reciprocally aligning polar active particles with repulsion, focusing on the performance on (and connection between) different levels of description. Linear stability analyses of the associated continuum model predict a profound influence of non-reciprocity, leading to phase separation, (anti-)flocking and asymmetric clustering behavior. On the microscopic level, particle simulations confirm the emergence of these dynamical phases and allow for a more in-depth investigation of (microscopic) properties, including orientational correlations and susceptibilities. The drastic impact of orientational couplings alone on the density dynamics is demonstrated in particle simulations without repulsion, where non-reciprocal alignment leads to the asymmetric formation of single-species polarized clumps. Overall, our findings demonstrate that certain dynamical properties, like a chase-and-run behavior in the asymmetrical clustering phase, are overlooked in mean-field continuum theory, making microscopic simulations an indispensable tool for studying the effects of non-reciprocal alignment couplings.
\end{abstract}

\maketitle


\section{\label{sec:introduction}Introduction}
Active matter systems are intrinsically out of equilibrium. Depending on the couplings between particles, these systems exhibit intriguing non-equilibrium transitions \cite{marchetti_simha_2013_hydrodynamics_soft_active_matter,bar_peruani_2020_self-propelled_rods,cates_tailleur_2015_mips} as well as pattern formation such as mesoscale turbulence \cite{wensink_yeomans_2012_mesoscale_turbulence,reinken_heidenreich_2018_derivation_mesoscale_turbulence}. Two paradigmatic transitions in active matter are motility-induced phase separation (MIPS) and flocking. MIPS emerges when the self-propulsion speed of particles depends on the local particle density, such that particles are slowed down in crowded situations and finally trapped \cite{cates_tailleur_2015_mips,OByrne_Zhao_2023_introduction_to_MIPS,Fily_Marchetti_2012_athermal_phase_separation_ABP,Redner_Baskaran_2013_structe_dynamics_phase-separating_active_fluid,Bialke_2013_microscopic_theory_phase_seperation,speck_2015_dynamical_mean_field_phase_separation}. This happens, e.g., for steric repulsion or quorum sensing. As shown in a multitude of experiments \cite{buttinoni_2013_dynamical_clustering,bauerle_2018_self-organization_quorum_sensing} and simulations \cite{Fily_Marchetti_2012_athermal_phase_separation_ABP,Redner_Baskaran_2013_structe_dynamics_phase-separating_active_fluid,levis_berthier_2014_clustering_monte_carlo_self-propelled_hard_disks}, the result of this trapping is phase separation into a dense state characterized by large clusters and a dilute state of essentially isolated particles. While MIPS can occur already in ``scalar'' active matter without orientational couplings, flocking requires the coupling of velocities of different particles, as it is common in ``polar'' active matter. In particular, velocity alignment of neighboring particles leads to the coherent ``flocking'' motion characterized by long-range polar order, as observed in the Vicsek model and its variants \cite{vicsek_1995_novel_type_phase_transition,bar_peruani_2020_self-propelled_rods,chate_raynaud_2008_collective_motion,chate_2020_dry_aligning_dilute_active_matter,toner_tu_2005_hydrodynamics_phases_of_flocks}, as well as in models of oppositely directed anti-flocking \cite{Menzel_2012_collective_motion_binary_self-propelled_particles,kreienkamp_klapp_2022_clustering_flocking_chiral_active_particles_non-reciprocal_couplings,Chatterjee_Noh_2023_flocking_two_unfriendly_species}.

The phase behavior becomes more complex in models involving both, translational and orientational couplings. It has been shown, e.g., that polar alignment of particles can stabilize the emergence of MIPS \cite{speck_2015_dynamical_mean_field_phase_separation,elena_2018_velocity_alignment_MIPS,elena_2021_phase_separation_self-propelled_disks} but also leads to arrested phase separation into smaller-sized clusters at long times \cite{Farrell_2012_Pattern_formation_self-propelled_particles_density-dependent_motility,van_der_linden_2019_interrupted_mips}. Not surprisingly, the interplay of positional and orientational couplings becomes even more delicate in chiral active matter \cite{levis_liebchen_2018_Brownian_chiral_disks,negi_maeda_2023_geometry-induced_dynamics_chiral_active_matter}, systems with non-separable (e.g., dipolar) interactions \cite{liao_2020_dynamical_self-assembly_dipolar_active_Brownian_particles} or mixtures \cite{kreienkamp_klapp_2022_clustering_flocking_chiral_active_particles_non-reciprocal_couplings,kreienkamp_klapp_PRL}.

The present paper addresses polar, repulsive active systems with an additional ingredient of complexity, that is, non-reciprocity.

Non-reciprocal couplings break the action-reaction symmetry and find applications in various real systems far from equilibrium \cite{You_Baskaran_Marchetti_2020_pnas,saha_scalar_active_mixtures_2020,fruchart_2021_non-reciprocal_phase_transitions,lavergne_bechinger_2019_group_formation_visual_perception-dependent_motility,Bowick_2022_symmetry_thermodynamcs_topology_active_matter,kreienkamp_klapp_2022_clustering_flocking_chiral_active_particles_non-reciprocal_couplings,Ivlev_2015_statistical_mechanics_where_newtons_third_law_is_broken,dinelli_tailleur_2023_non-reciprocity_across_scales,frohoff_thiele_2023_nonreciprocal_cahn-hilliard}. Examples include biologically motivated modeling of animal behavior \cite{haluts_gov_2024_active_particle_models_non-reciprocity,basaran_kocabas_2024_parity-time_symmetry_breaking_bacteria} and agents with vision cones \cite{lavergne_bechinger_2019_group_formation_visual_perception-dependent_motility,barberis_peruani_2016_minimal_cognitive_flocking_model,loos_martynec_2023_long-range_order_non-reciprocal_XY_model,avni_vitelli_2023_non-reciprocal_Ising_model}, predator-prey systems \cite{lotka_1920_lotka_volterra_model,volterra_1926_lotka_volterra_model,mobilia_taeuber_2007_phase_transitions_lotka-volterra,tsyganov_biktashev_2003_quasisoliton_predator-prey_system}, multi-component bacterial suspensions \cite{theveneau_mayor_2013_chase-and-run_cells,xiong_tsimring_2020_flower-like_patterns_multi-species_bacteria}, quorum sensing \cite{duan_mahault_2023_dynamical_pattern_formation_without_self-attraction,knevzevic_stark_2022_collective_motion_non-reciprocal_orientational_interactions}, and phoretic interactions of colloids \cite{agudo-canalejo_golestanian_2019_active_phase_separation_chemically_interacting_particles,soto_2014_self-assembly_catalytically_active_colloidal_molecules,saha_golestanian_2019_pairing_waltzing_scattering_chemotactic_active_colloids}. Generally, non-reciprocity can occur when interactions are mediated by a non-equilibrium environment as, for instance, in odd solids \cite{scheibner_vitelli_2020_odd_elasticity,fruchart_vitelli_2023_odd_viscosity_and_elasticity} or by hydrodynamic interactions \cite{maity_morin_2023_spontaneous_demixing_binary_colloidal_flocks,banerjee_rao_2022_nonreciprocity_compressible_viscoelastic_fluid,gupta_ramaswamy_2022_active_nonreciprocal_attraction_elastic_medium}.

Over the last years, it has been established that non-reciprocity can have drastic effects on the observed collective dynamics, including the spontaneous formation of traveling states in diffusive systems with conserved scalar fields \cite{You_Baskaran_Marchetti_2020_pnas,saha_scalar_active_mixtures_2020}, or the emergence of chiral motion in \mbox{(anti-)}aligning polar systems \cite{fruchart_2021_non-reciprocal_phase_transitions} via so-called exceptional points. Further field theoretical studies \cite{duan_mahault_2023_dynamical_pattern_formation_without_self-attraction,brauns_marchetti_2023_non-reciprocal_pattern_formation_of_conserved_fields,Frohoff-Huelsmann_Thiele_2021_cahn-hilliard_nonvariational_coupling,alston_bertrand_2023_irreversibility_non-reciprocal_PT-breaking,Martin_Vitelli_2023_exact_model_collective_motion_non-reciprocal_active_matter} as well as particle simulations \cite{duan_mahault_2023_dynamical_pattern_formation_without_self-attraction,alston_bertrand_2023_irreversibility_non-reciprocal_PT-breaking,mandal_sollich_2022_robustness_travelling_states_generic_non-reciprocal_mixture,chiu_Omar_2023_phase_coexistence_implications_violating_Newtons_third_law} have elucidated the emergence of a variety of dynamical phases depending on the precise non-reciprocal interactions between particles. Yet, so far, non-reciprocity has been studied rather separately in systems with either purely translational (i.e., position-dependent) or purely orientational interactions, neglecting density dynamics.

Motivated by this gap, we have recently studied the effect of non-reciprocal alignment couplings on density dynamics. To this end, we considered a minimal two-component model combing non-trivial density dynamics (induced by repulsive interactions) and non-reciprocal polar couplings \cite{kreienkamp_klapp_PRL}. Using a mean-field stability analysis and particle simulations, we have unraveled an unexpected phenomenon induced by non-reciprocity, namely, asymmetric clustering in a system with fully symmetric translational interactions.

While \cite{kreienkamp_klapp_PRL} primarily addresses the clustering behavior from a continuum perspective, we here provide a comprehensive particle-based characterization of dynamical and structural properties of all the emerging non-equilibrium phases, including aspects of orientational ordering. We further present particle simulation results without repulsion, which emphasize the significant effect of non-reciprocal alignment on density dynamics. By performing this systematic particle-based analysis we aim at exploring the validity, but also the limits, of our mean-field continuum description \cite{marchetti_simha_2013_hydrodynamics_soft_active_matter,dinelli_tailleur_2023_non-reciprocity_across_scales}. For specific non-reciprocal models of active matter, exact coarse-grained equations have already been derived, which can accurately predict the transition to collective motion \cite{Martin_Vitelli_2023_exact_model_collective_motion_non-reciprocal_active_matter}. However, such an exact coarse-graining is impossible for the present model, which includes non-linear interactions. Thus, a particle-based investigation is indispensable for an overall understanding of the collective dynamics. Indeed, while our mean-field approach turns out to predict the main features, certain dynamical properties, like chase-and-run behavior, are overlooked.

The paper is organized as follows. In Sec.~\ref{sec:model}, we start with an introduction to the microscopic model. Our methods of analysis are introduced in Sec.~\ref{sec:methods_of_analysis}. In Sec.~\ref{sec:phase_behavior_stability_diagram} we present our results for the non-equilibrium phase behavior from particle simulations and linear stability analyses.
In Sec.~\ref{sec:correlation_functions}, we show how characteristics of different dynamical phases are reflected by pair correlation functions, which are subsequently used for a systematic fluctuation analysis in terms of structure factors in Sec.~\ref{sec:structure_factors}. We close with a discussion of our results in Sec.~\ref{sec:summary_conclusion}. The paper includes several appendices addressing primarily the continuum model and its relation to the particle-based model as well as the linear stability analysis.

\section{\label{sec:model}Model}

\subsection{\label{sec:microscopic_model}Microscopic model}
We consider a two-dimensional system of circular active particles comprising two species $a=A,B$. The binary mixture contains $N = N_A + N_B$ particles, which are located at positions $\bm{r}_{\alpha}$ (with $\alpha=i_a = 1,...,N_a$) and move like active Brownian particles (ABP) \cite{romanczuk_schimansky-geier_2012_active_brownian_particles} with additional torque due to orientational couplings. The particles are confined to a box of size $L\times L$, which is subject to periodic boundary conditions. They self-propel with velocity $v_0$ in the direction given by $\bm{p}_{\alpha}(t)=(\cos\,\theta_{\alpha}, \sin\,\theta_{\alpha})^{\rm{T}}$, where $\theta_{\alpha}$ is the polar angle. The dynamics are governed by the set of the overdamped Langevin equations (LE)
\begin{subequations} \label{eq:Langevin_eq}
	\begin{align}
		\dot{\bm{r}}_{\alpha}(t) &= v_0\,\bm{p}_{\alpha}(t) + \mu_{r} \sum_{\beta\neq\alpha} \bm{F}_{\rm{rep}}^{\alpha}(\bm{r}_{\alpha},\bm{r}_{\beta}) + \bm{\xi}_{\alpha}(t) \label{eq:Langevin_r}\\
		\dot{\theta}_{\alpha}(t) &= \mu_{\theta} \sum_{\beta\neq\alpha} \mathcal{T}_{\rm al}^{\alpha}(\bm{r}_{\alpha},\bm{r}_{\beta},\theta_{\alpha},\theta_{\beta}) + \eta_{\alpha}(t) \label{eq:Langevin_theta},
	\end{align}
\end{subequations}
where the sums over particles $\beta=j_b=1,...,N_b$ couple the dynamics of particle $\alpha$ to the positions and orientations of all other particles of both species, $b=A,B$. 

The translational LE \eqref{eq:Langevin_r} involves the repulsive force $\bm{F}_{\rm{rep}}^{\alpha} = - \sum_{\beta\neq\alpha} \nabla_{\alpha} U(r_{\alpha\beta})$ between hard disks, derived from the Weeks-Chandler-Andersen (WCA) potential \cite{weeks_1971_Weeks-Chandler-Andersen_potential} 
{\small
	\begin{equation}
		\label{eq:WCA_potential}
		U(r_{\alpha\beta}) = \begin{cases}
			4\epsilon \left[\left( \sigma/r_{\alpha\beta}\right)^{12} - \left( \sigma/r_{\alpha\beta}\right)^6 + \frac{1}{4} \right], \, {\rm if} \, r_{\alpha\beta}<r_{\rm c}\\
			0, \ {\rm else} ,
		\end{cases}
	\end{equation}
}%
where $r_{\alpha\beta} = \vert \bm{r}_{\alpha\beta}\vert = \vert \bm{r}_{\alpha}-\bm{r}_{\beta} \vert$. The characteristic energy scale of the potential is given by $\epsilon$. The cut-off distance is $r_{\rm c }=2^{1/6}\,\sigma$. We take the particle diameter $\sigma$ as characteristic length scale, $\ell = \sigma$. 

The rotational LE \eqref{eq:Langevin_theta} contains the torque
\begin{equation}
	\label{eq:torque}
	\mathcal{T}_{\rm al}^{\alpha}(\bm{r}_{\alpha}, \bm{r}_{\beta}, \theta_{\alpha}, \theta_{\beta}) = k_{ab}\, \sin(\theta_{\beta}-\theta_{\alpha}) \, \Theta(R_{\theta}-r_{\alpha\beta})
\end{equation}
of strength $k_{ab}$. The constants $k_{ab}$ can be positive or negative. Further, $\Theta(R_{\theta}-r_{\alpha\beta})$ is the step function with $\Theta(R_{\theta}-r_{\alpha\beta})= 1$, if $r_{\alpha\beta} < R_{\theta}$, and zero otherwise. From Eq.~\eqref{eq:torque} it follows that particles of species $a$ tend to orient parallel, i.e., align with neighboring particles of species $b$ when $k_{ab}>0$. They orient anti-parallel, i.e., anti-align when $k_{ab}<0$. We assume equal intraspecies alignment for the two species, i.e., $k_{AA}=k_{BB}>0$. The interspecies couplings can be reciprocal or non-reciprocal. For \textit{reciprocal} couplings defined by the choice $k_{AB} = k_{BA}$, particles of species $A$ align (or anti-align) with particles of species $B$ in the same way as particles of species $B$ with particles of species $A$. Here, we specifically allow for \textit{non-reciprocal} orientational couplings, where $k_{AB}\neq k_{BA}$. Note that these orientational couplings are separable, i.e.,~independent of the spatial configuration (unlike, e.g., hydrodynamic or dipolar interactions).

As seen from LEs \eqref{eq:Langevin_eq}, both the positions and orientations of the particles are subject to thermal noise, modeled as Gaussian white noise processes $\bm{\xi}_{\alpha}$(t) and $\eta_{\alpha}(t)$ of zero mean and variances $\langle \xi_{\alpha,k}(t) \xi_{\beta,l}(t') \rangle = 2\,\xi\,\delta_{\alpha\beta}\,\delta_{kl} \,\delta(t-t')$ and $\langle \eta_{\alpha}(t) \eta_{\beta}(t') \rangle = 2\,\eta\,\delta_{\alpha\beta}\,\delta(t-t')$, respectively. The (Brownian) time a (passive) particle needs to travel over its own distance is $\tau = \sigma^2/\xi$, which we take as characteristic time scale. The mobilities are assumed to fulfill the Einstein relation and are connected to thermal noise via $\mu_{r} = \beta\,\xi$ and $\mu_{\theta} = \beta \, \eta$, where $\beta^{-1}=k_{\rm B}\,T$ is the thermal energy with Boltzmann's constant $k_{\rm B}$ and temperature $T$. 

Thus, we choose all particles to be equal with the same steric interactions between all types of particles. The two species only differ in their alignment couplings with respect to each other (i.e., $g_{AB}$ and $g_{BA}$).

To study the emerging dynamical structures in our system, we perform numerical Brownian Dynamics (BD) simulations of the LEs \eqref{eq:Langevin_eq}. To this end, we introduce the following dimensionless parameters:
the average area fraction $\Phi_a$ of species $a$,
\begin{equation}
	\Phi_a = \frac{N_a\,\pi\,\ell^2}{4\,L^2} = \rho_0^a\,\frac{\pi\,\ell^2}{4} ,
\end{equation}
with (number) density $\rho_0^a = N_a/L^2$, the reduced orientational coupling parameter,
\begin{equation}
	g_{ab} = k_{ab}\,\mu_{\theta}\,\tau,
\end{equation}
and the P\'eclet number,
\begin{equation}
	{\rm Pe} = \frac{v_0\,\tau}{\ell} ,
\end{equation}
which quantifies the persistence of the motion of particles. We perform simulations at a fixed combined average area fraction $\Phi = 0.4$, where $\Phi_A=\Phi_B=0.2$, and P\'eclet number ${\rm Pe}=40$.  Further, we set $g_{AA}=g_{BB}=3$, while varying the orientational couplings strengths $g_{AB}$ and $g_{BA}$. The parameters are chosen in such a way that the non-aligning system ($g_{ab}=0 \, \forall \, a,b$) exhibits MIPS. The choice of $g_{AA}=g_{BB}=3$ ensures that flocking only emerges for large products $g_{AB}\,g_{BA}>0$ and is suppressed for smaller $g_{AB}\,g_{BA}$ by rotational diffusion (see \cite{kreienkamp_klapp_PRL}). The BD simulations are performed with $N=5000$ particles, with equal particle numbers $N_A = N_B = 2500$ of both species. The repulsive strength is chosen to be $\epsilon^{*} = \epsilon / (k_{\rm B}\,T) = 100$, where the thermal energy is set to be the energy unit ($k_{\rm B}\,T=1$). The translational and rotational diffusion constants are then given by $\xi = 1 \, \ell^2/\tau$ and $\eta = 3 \cdot 2^{-1/3}/\tau$, respectively. The cut-off distance for the torque is chosen to be $R_{\theta}=2\,\ell$. The simulations are performed by initializing the system in a random configuration. We then integrate the equations of motions using an Euler-Mayurama algorithm \cite{kloeden_platen_2011_numerical_solution_SDE}, and let the system evolve into a non-equilibrium steady state before measuring quantities for phase characterization. Indeed, we have found saturation of the dynamics into non-equilibrium steady state at all parameters considered. We employ a timestep of $\Delta t = 10^{-5}\,\tau$. Typical simulations last for $120\,\tau$.

\subsection{Continuum model}
A central goal of this paper is to compare our particle-based simulation results with predictions from a mean-field continuum theory. The continuum model is derived in \mbox{Appendix \ref{app:continuum_model}}, where we follow essentially the steps in \cite{kreienkamp_klapp_2022_clustering_flocking_chiral_active_particles_non-reciprocal_couplings}.
The continuum model describes the spatio-temporal evolution of the particle densities $\rho^a(\bm{r},t)$ ($a=A,B$) via the continuity equation
\begin{equation}
	\label{eq:continuum_eq_density_main}
	\partial_t \rho^a +  \nabla \cdot \big[v^{\rm eff}(\rho)\, \bm{w}^a -  D_{\rm t}\,\nabla\,\rho^a\big] = 0 ,
\end{equation}
where $v^{\rm eff}(\rho) = {\rm Pe} - z\,\rho$ with $\rho = \rho^A + \rho^B$ denotes the effective velocity reduction of particles in high-density regimes, see Appendix \ref{sapp:steric_effects_continuum}.
The polarization densities $\bm{w}^a(\bm{r},t)$ measure the overall orientation of particles at a certain position via $\bm{w}^a/\rho^a$. They evolve according to
\begin{equation}
	\label{eq:continuum_eq_polarization_main}
		\begin{split}
			\partial_t \bm{w}^a 
			=& - \frac{1}{2} \, \nabla \,\big(v^{\rm eff}(\rho)\, \rho^a\big)  - D_{\rm r }\, \bm{w}^{a} + \sum_b g'_{ab} \, \rho^a\, \bm{w}^b \\
			&+  D_{\rm t}\,\nabla^2\,\bm{w}^a + \frac{v^{\rm eff}(\rho)}{16\,D_{\rm r}} \, \nabla^2\,\Big(v^{\rm eff}(\rho)\,\bm{w}^a \Big) \\
			&- \sum_{b,c} \frac{ g'_{ab}\,g'_{ac}}{2\,D_{\rm r}} \, \bm{w}^a \, (\bm{w}^b \cdot \bm{w}^c) \\
			& + O(\bm{w}\nabla \bm{w}) + O(\nabla \rho \nabla \bm{w}) .
		\end{split}
\end{equation}
These equations are non-dimensionalized and scaled with particle density $\rho_0^a$. Translational and rotational diffusion constants are $D_{\rm t}$ and $D_{\rm r}$. On the continuum level, the relative orientational coupling parameter is given by $g'_{ab} = g_{ab} \, R_{\theta}^2 \, \rho_0^b/2$ and scales with the average density $\rho_0^b$. The full expressions are given in Eqs.~\eqref{eq:continuum_eq_density}-\eqref{eq:continuum_eq_polarization} in \mbox{Appendix \ref{app:continuum_model}}.
The choice of parameters in the continuum model is summarized in Appendix \ref{sapp:parameter_choice_continuum}. \cor{In Appendix \ref{sapp:steric_effects_continuum}, we additionally briefly discuss the limiting case of passive systems with $v^{\rm eff}=0$.} We use the continuum equations as a starting point of a mean-field stability analysis as described in Appendix \ref{app:linear_stability_analysis}.

\section{Methods of analysis}
\label{sec:methods_of_analysis}
In this section, we introduce the quantities and methods that will be used to characterize the dynamical behavior of the system on the particle level. 
The classification of the observed non-equilibrium states in terms of these quantities is summarized in Table \ref{tab:target_quantities_phase_characterization}.

\begin{table*}
	\begin{tabular}{p{4cm}p{4.5cm}p{3.5cm}p{3cm}}
		\hline
		Non-equilibrium phase & Polarization & Largest cluster size & Distribution of \newline local area fraction\\ 
		\hline
		\hline
		phase separation & $P_{\rm combi}=P_A=P_B=0\, (\leq 0.6)$ & $\mathcal{N}_{{\rm lcl}} \approx 1\, (\geq 0.3)$ \newline $\mathcal{N}_{{\rm lcl},a} \approx 0 \, (\leq 0.1)$ & double peak \\ 
		\hline
		flocking (\& phase sep.)& $P_{\rm combi}=P_A=P_B=1\,(\geq 0.6)$ & $\mathcal{N}_{{\rm lcl}} \approx 1\, (\geq 0.3)$ \newline $\mathcal{N}_{{\rm lcl},a} \approx 0 \, (\leq 0.1)$ & double peak \\  
		\hline
		anti-flocking & $P_{\rm combi}=0\, (\leq 0.3)$, \newline $P_A=P_B=1\, (\geq 0.6)$  & $\mathcal{N}_{{\rm lcl}} \approx \mathcal{N}_{{\rm lcl},a} \approx 1\, (\geq 0.6)$ & double peak \\ 
		\hline
		disorder & $P_{\rm combi}=P_A=P_B=0\, (\leq 0.3)$ & $\mathcal{N}_{{\rm lcl}} \approx \mathcal{N}_{{\rm lcl},a} \approx 0\, (\leq 0.3)$ & single peak \\ 
		\hline
		asymmetric clustering & $P_{\rm combi}\approx P_a \gg P_b$, $a\neq b$ & $\mathcal{N}_{{\rm lcl}} \approx \mathcal{N}_{{\rm lcl},a} \gg \mathcal{N}_{{\rm lcl},b}$,\newline $a\neq b$ &  \\ 
		\hline
	\end{tabular}
	\caption{\label{tab:target_quantities_phase_characterization}Characterization of the non-equilibrium phases in terms of polarization, largest cluster size, and local area fraction in the binary mixture with $N_A=N_B$. Values in parenthesis denote the cut-off values used to characterize the emerging phases in our particle simulations.}
\end{table*}

\subsection{\label{ssec:methods_of_analysis_clustering_behavior}Clustering behavior}
Cluster formation is a well-studied phenomenon in one-component, non-polar ABP systems. It occurs at sufficiently high density and particle motility, eventually leading to phase separation into dilute and dense regions \cite{cates_tailleur_2015_mips,buttinoni_2013_dynamical_clustering,Bialke_2013_microscopic_theory_phase_seperation,van_der_linden_2019_interrupted_mips,bauerle_2018_self-organization_quorum_sensing,liu_2019_self-driven_phase_transitions,OByrne_Zhao_2023_introduction_to_MIPS}. Also in our system, phase separated states can emerge due to the repulsive interactions between particles.

To determine whether the system is in a phase-separated state, we calculate the position-resolved local area fraction $\overline{\Phi}(x,y)$. For details regarding the calculation, see Appendix \ref{app:analysis_methods_details_clustering}. The distribution $P(\overline{\Phi})$ exhibits a double-peak structure when dense and dilute regions coexist \cite{blaschke_2016_phase_separation,Liao_2018_circle_swimmers_monolayer,liao_2020_dynamical_self-assembly_dipolar_active_Brownian_particles}. Considering separately the local area distributions of $A$- and $B$-particles gives us, in addition, information on spatially inhomogeneous compositions.

To characterize quantitatively the clusters within the system, we determine the largest cluster size in terms of a time-average of the ratio of all particles in the largest cluster, given as $\mathcal{N}_{\rm lcl} = \langle n_{\rm lcl} \rangle_t/N$. We also distinguish between clusters made of particles of any species and clusters made of particles of only one species $A$ or $B$, see Appendix \ref{app:analysis_methods_details_clustering}.

The clustering behavior is analyzed via averages in the non-equilibrium steady state. To this end, we perform time averages between $70$ and $120\,\tau$ after initialization of a disordered state.

\subsection{Global orientational ordering}
In systems with sufficiently large alignment strengths between particles, 
a phase transition from an isotropic disordered state to an oriented flocking state can occur. This flocking state is characterized by a non-zero value of the global polarization, see, e.g., \cite{martin_gomez_pagonabarraga_2018_collective_motion_ABP_alignment_volume_exclusion}. In our binary mixture, we define the polarization $P_a$ of particles belonging to species $a$ as
\begin{equation}
	\label{eq:polarization}
	P_a(t) = \vert \bm{P}_a(t) \vert = \Bigg\vert \frac{1}{N_a} \sum_{a_i=1}^{N_a}  \bm{p}_{a_i}(t)  \Bigg\vert.
\end{equation}
The quantity $P_{\rm combi} = \vert \bm{P}_A + \bm{P}_B \vert$ then measures the polarization of the entire system involving both species $A$ and $B$. In a perfect flocking state, where all particles move coherently in the same direction, the polarization order parameters $P_{\rm combi}$, $P_A$ and $P_B$ take the value of unity. In the anti-flocking state, $P_A=P_B=1$ but $P_{\rm combi}=0$. When the system is disordered, particle orientations are uncorrelated, yielding $P_{\rm combi}=P_A=P_B=0$. Where appropriate, we have also studied the polarization of particles within a single cluster.

As in equilibrium phase transitions, it is furthermore useful to investigate the susceptibility of the polarization order parameter $P$. Close to the flocking transition, one expects the susceptibility to peak \cite{Baglietto_Albano_2008_Finite-size_scaling_analysis_self-driven_indiviudals}. Here, we consider the susceptibility $\chi_{\rm combi}(P_{\rm combi})$ of the whole system as well as the single-species susceptibilities $\chi_a(P_a)$. They can be calculated from the fluctuations in their respective order parameters $P_{\rm combi}$ and $P_a$ \cite{Baglietto_Albano_2008_Finite-size_scaling_analysis_self-driven_indiviudals,Adhikary_Santra_2022_pattern_formation_phase_transition_binary_mixture,martin_gomez_pagonabarraga_2018_collective_motion_ABP_alignment_volume_exclusion}. Specifically, we determine the susceptibilities as 
\begin{equation} \label{eq:susceptibility_scalar_polarization_definition_combi}
	\begin{split}
		\chi_{\rm combi}&(P_{\rm combi}) = N\,{\rm Var}(P_{\rm combi}) \,  \\
		&= \frac{1}{N} \Big ( \Big \langle \Big \vert \sum_{i=1}^N \bm{p}_i \Big \vert \, \Big \vert \sum_{j=1}^N \bm{p}_j \Big \vert \Big \rangle  - \Big \langle \Big \vert \sum_{i=1}^N \bm{p}_i \Big \vert \Big \rangle^2 \Big )
	\end{split}
\end{equation}
and
\begin{equation} \label{eq:susceptibility_scalar_polarization_definition_single-species}
	\begin{split}
		&\chi_{a}(P_a) = N_a\,{\rm Var}(P_{a}) \\
		& \ \ = \frac{1}{N_a} \Big ( \Big \langle \Big \vert \sum_{a_i=1}^{N_a} \bm{p}_{a_i} \Big \vert \, \Big \vert \sum_{a_j=1}^{N_a} \bm{p}_{a_j} \Big \vert \Big \rangle - \Big \langle \Big \vert \sum_{a_i=1}^{N_a} \bm{p}_{a_i} \Big \vert \Big \rangle^2 \Big )
	\end{split}
\end{equation}
where ${\rm Var}(P) = \langle P^2 \rangle - \langle P \rangle^2$ is the variance, and the angular brackets denote ensemble averages. For technical details regarding the averaging, see end of Sec.~\ref{ssec:methods_of_analysis_clustering_behavior}.

\subsection{Pair correlation function}
Information on the translational structure in our active binary mixture is captured by the pair correlation function $G_{ab}(\bm{r})$. The function describes the distribution of distance vectors $\bm{r}$ between pairs of particles belonging to species $a$ and $b$ under assumption of spatial homogeneity \cite{Bartnick_Loewen_2015_structural_correlations_in_nonreciprocal_systems}. In two dimensions, the distance vector can be parametrized as $\bm{r} = (r,\phi)$ with $r=\vert\bm{r}\vert$ and angle $\phi=\sphericalangle(\bm{r}, \bm{p})$ between the relative position $\bm{r}$ and self-propulsion direction $\bm{p}$. In passive systems, one expects $G_{ab}$ to depend on $r$ alone. However, in active systems close to motility-induced phase separation, the probability distribution depends not only on the distance $r$ but also on the spatial configuration of two particles, and thus, $\phi$ \cite{Bialke_2013_microscopic_theory_phase_seperation,buttinoni_2013_dynamical_clustering,elena_2021_phase_separation_self-propelled_disks}. This anisotropy arises from the blocking of particles in the direction of self-propulsion, leading to a density-dependent reduction of self-propulsion speed as described in more detail in Appendix \ref{sapp:steric_effects_continuum}. Here, we define $G_{ab}(\bm{r})$ as \cite{hansen_mcDonald_2013_theory_simple_liquids,Bartnick_Loewen_2015_structural_correlations_in_nonreciprocal_systems}
\begin{equation}
	\label{eq:pair_distribution_function}
	G_{ab}(\bm{r}) = \frac{1}{\Omega} \, \sum_{a_i=1}^{N_{a}} \sum_{\substack{b_j=1\\ (b_j\neq a_i)}}^{N_{b}} \Big \langle \delta(\bm{r} - (\bm{r}_{a_i} - \bm{r}_{b_j})) \Big \rangle,
\end{equation}
where $\Omega = N_a\,N_b/V$ is the normalization and $V=L^2$ represents the volume of the system. $G_{ab}(\bm{r})$ tends to unity for $r\rightarrow \infty$ and vanishes for $r \rightarrow 0$ due to steric repulsion between particles. Numerically, we determine $G_{ab}(r,\phi)$ by counting the particles found in small area fractions of distance $r+\Delta r$ and angle $\phi + \Delta \phi$ from the reference particle, such that we additionally normalize with the area fraction element $\Delta A = r\,\Delta r \, \Delta\phi$, leading to 
\begin{equation}
	G_{ab}(r,\phi) = \frac{1}{\Omega_{\rm n}} \, \sum_{a_i=1}^{N_{a}} \sum_{\substack{b_j=1\\ (b_j\neq a_i)}}^{N_{b}} \Big \langle \delta(r^{ab}_{ij} - r)\,\delta({\phi^{ab}_{ij}-\phi}) \Big \rangle
\end{equation}
with $\Omega_{\rm n} = N_{a}\,N_{b}\,\Delta A/V$. The particle distance and relative angle are calculated as $r^{ab}_{ij} = \vert \bm{r}_{b_j} - \bm{r}_{a_i} \vert$ and \mbox{$\phi_{ij}^{ab} = \sphericalangle(\bm{r}_{b_j}-\bm{r}_{a_i}, \bm{p}_{a_i})$}.

To account, in addition, for the orientational correlations of the particles, we extent the definition of the pair distribution function $G_{ab}(\bm{r})$ to the correlation function $C_{ab}(\bm{r})$ defined as \cite{cavagna_grigera_2018_physics_of_flocking_correlations}
{\small
\begin{equation}
		\label{eq:orientational_correlation}
		C_{ab}(\bm{r}) = \frac{1}{\Omega} \, \sum_{a_i=1}^{N_{a}} \sum_{\substack{b_j=1\\ (b_j\neq a_i)}}^{N_{b}} \Big \langle \bm{p}_{a_i} \cdot \bm{p}_{b_j} \, \delta(\bm{r} - (\bm{r}_{a_i} - \bm{r}_{b_j})) \Big \rangle,
\end{equation}}%
where $\bm{p}_{a_i}$ and $\bm{p}_{b_j}$ represent the orientation vectors of particles $i$ and $j$ belonging to species $a$ and $b$, respectively. By definition, $C_{ab}(\bm{r})$ carries both, translational and orientational information. In fully aligned systems, where $\langle \bm{p}_{a_i} \cdot \bm{p}_{b_j} \rangle = 1 \ \forall \ i,j {\rm \ and \ } a,b$, the orientational correlation function reduces to the common pair distribution function $G_{ab}(\bm{r})$, exhibiting positive peaks at typical particle distances and approaching unity in the long-distance limit. Conversely, in anti-aligned systems with $\langle \bm{p}_{a_i} \cdot \bm{p}_{b_j} \rangle = -1\ \forall \ i,j {\rm \ and \ } a\neq b$, $C_{ab}(\bm{r}) = -G_{ab}(\bm{r})$. Fully disordered systems with $\langle \bm{p}_{a_i} \cdot \bm{p}_{b_j} \rangle = 0\ \forall \ i,j {\rm \ and \ } a,b$ have an orientational correlation function of zero. Finally, for $r\rightarrow \infty$, $C_{ab}(\bm{r}) \rightarrow \bm{P}_a \cdot \bm{P}_b$, that is, the product of orientational order parameters.

Note that, by construction, the pair distribution function as well as orientational correlation function are symmetric in the sense that $G_{AB} = G_{BA}$ and $C_{AB} = C_{BA}$. The effect of non-reciprocal interspecies couplings is expected to manifest itself in differences between the single-species correlations, such that $G_{AA} \neq G_{BB}$ and $C_{AA} \neq C_{BB}$ \cite{Bartnick_Loewen_2015_structural_correlations_in_nonreciprocal_systems}.

\begin{figure*}
	\includegraphics[width=\linewidth]{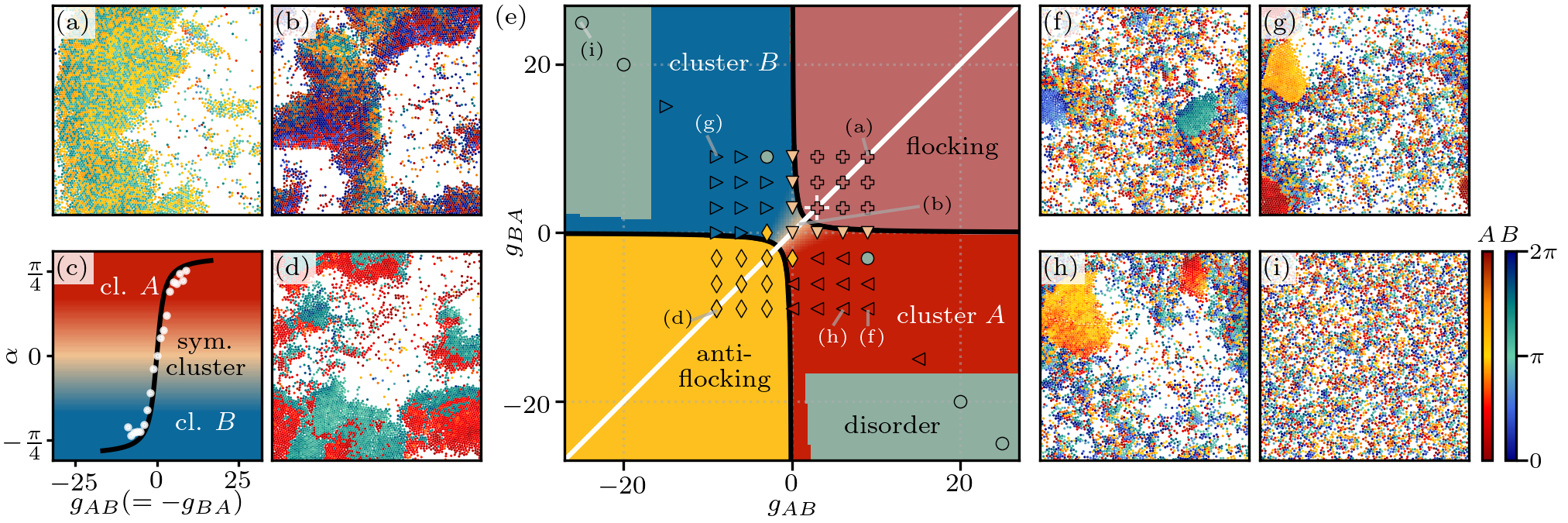}
	\caption{\label{fig:stability_diagram_with_snapshots}Stability diagram and particle simulation snapshots. The stability diagram (e) is obtained from a linear stability analysis of the disordered base state in the continuum Eqs.~\eqref{eq:continuum_eq_density_main} and \eqref{eq:continuum_eq_polarization_main}. Marker points represent color-coded dynamical phases from particle simulations, characterized as summarized in Table \ref{tab:target_quantities_phase_characterization}. (c) Mean-field predictions (line) for angle $\alpha$ characterizing asymmetric clustering. Dots are from structure factor analysis (see Fig.~\ref{fig:structure_factor_matrix_eigenvalues}). Snapshots (a,b,d,f-i) show particle simulations ($t=100\,\tau$) with (a) $g_{AB}=g_{BA}=9$, (b) $g_{AB}=g_{BA}=1$, (d) $g_{AB}=g_{BA}=-9$, (f) $g_{AB}=-g_{BA}=-9$, (g) $g_{AB}=-g_{BA}=9$, (h) $g_{AB}=6$, $g_{BA}=-9$, and (i) $g_{AB}=-g_{BA}=-25$. Other parameters are $g_{AA}=g_{BB}=3$, ${\rm Pe}=40$, $\Phi=0.4$, $z=14.8$, $D_{\rm t}=9$, and $D_{\rm r} = 3 \cdot 2^{-1/3}$.}
\end{figure*}

\subsection{Structure factor matrix in binary mixture}
\label{ssec:analysis_methods_structure_factor_matrix}
To characterize the density fluctuations close to phase transitions in our binary mixture, we took inspiration from established procedures developed for equilibrium mixtures \cite{chen_forstmann_1992_demixing_binary_Yukawa_fluid, chen_forstmann_1992_phase_instability_fluid_mixtures}. Specifically, our approach involves the computation of density fluctuations of the form $\langle \delta \rho_{a}(\bm{r}) \, \delta \rho_{b}(\bm{r}') \rangle$, where $a,b = A,B$ and $\delta \rho_{a}(\bm{r}) = \rho_{a}(\bm{r}) - \rho_{0}^a$ with $\rho_{0}^a$ as the density of the homogeneous system. Here, we only consider instantaneous fluctuations and neglect all time-dependencies. Using the definition of the particle density in terms of sums over $\delta$-functions, that is, $\rho_{a}(\bm{r}) = \sum_{a_i=1}^{N_{a}} \delta(\bm{r}-\bm{r}_{a_i})$, one finds
\begin{equation}
	\langle \delta \rho_{a}(\bm{r}) \, \delta \rho_{b}(\bm{r}') \rangle = \langle \rho_{a}(\bm{r}) \, \rho_{b}(\bm{r}') \rangle - \rho_{0}^a\,\rho_{0}^b ,
\end{equation}
where $\langle \cdot \rangle$ denotes the ensemble average and $\rho_0^a = \langle \rho_a(\bm{r}) \rangle$ in a uniform system.
Further, the density correlation $\langle \rho_{a}(\bm{r}) \, \rho_{b}(\bm{r}') \rangle$ is linked to the two-particle density
\begin{equation}
	\rho^{(2)}_{ab}(\bm{r},\bm{r}') = \Big \langle \sum_{a_i = 1}^{N_{a}} \sum_{\substack{b_j = 1 \\ b_j \neq a_i}}^{N_{b}} \delta(\bm{r}-\bm{r}_{a_i}) \, \delta(\bm{r}'-\bm{r}_{b_j})  \Big \rangle
\end{equation}
via \cite{landau_lifshitz_2013_statistical_physics}
\begin{equation}
	\langle \rho_{a}(\bm{r}) \, \rho_{b}(\bm{r}') \rangle = \rho^{(2)}_{ab}(\bm{r},\bm{r}') + \delta_{ab}\, \rho_{0}^a\,\delta(\bm{r}-\bm{r}') .
\end{equation}
In homogeneous systems, the two-particle density depends only on the difference $\bm{r}-\bm{r}'$ and is, in turn, related to the pair correlation function $G_{ab}(\bm{r}-\bm{r}')$ via \cite{hansen_mcDonald_2013_theory_simple_liquids}
\begin{equation}
	\rho_{ab}^{(2)}(\bm{r}-\bm{r}') = G_{ab}(\bm{r}-\bm{r}') \, \rho_{0}^a\,\rho_{0}^b.
\end{equation}
This relationship allows the calculation of density fluctuations using the pair correlation function:
\begin{equation}
	\begin{split}
		\langle \delta &\rho_{a}(\bm{r}) \, \delta \rho_{b}(\bm{r}') \rangle \\
		& = \rho_{0}^a\,\rho_{0}^b \, h_{ab}(\bm{r}-\bm{r}') + \delta_{ab} \, \rho_{0}^a \, \delta(\bm{r}-\bm{r}') ,
	\end{split}
\end{equation}
where $h_{ab}(\bm{r}-\bm{r}')=G_{ab}(\bm{r}-\bm{r}') -1$ denotes the total correlation function.
In Fourier space, and neglecting angle dependencies, that is, $h_{ab}(\bm{r}-\bm{r}')=h_{ab}(\vert\bm{r}-\bm{r}'\vert)$, the calculation simplifies to
\begin{equation}
	\label{eq:mean_fluctuation_products}
	\frac{1}{V}\,\langle \delta \hat{\rho}_{a}(\bm{k}) \, \delta \hat{\rho}_{b}(-\bm{k}) \rangle = \rho_{0}^a \,\rho_{0}^b\, \hat{h}_{ab}(k) + \delta_{ab} \, \rho_{0}^a,
\end{equation}
where Fourier transformed quantities are indicated by a hat, $\hat{\cdot}$ \footnote{We define the (two-dimensional) Fourier transform as $\hat{f}(\bm{k}) =  \int_{\mathbb{R}^2} f(\bm{r}) \, {\rm e}^{-2\pi i \bm{r}\cdot \bm{k}} {\rm d}\bm{r}$ (with inverse $f(\bm{r}) =  \int_{\mathbb{R}^2} \hat{f}(\bm{k}) \, {\rm e}^{2 \pi i \bm{r}\cdot \bm{k}} {\rm d}\bm{k}$). For a radially symmetric integral kernel, the two-dimensional Fourier transform is a Hankel or Fourier-Bessel transform (of order zero) with $\hat{f}(k) = 2\,\pi \int_{0}^{\infty} f(r) \, J_0(2\,\pi\,k\,r)\,r\,{\rm d}r$, where $J_0(z)$ is the zeroth order Bessel function of the first kind.} and $\vert \bm{k} \vert = k$. We note already here that in the present system, the assumption of homogeneity and isotropy holds only for short times after starting from a random configuration.

To characterize the type of phase transition related to densities within the binary mixture, we consider two different types of fluctuations: fluctuations in the total density, $\delta\hat{\rho}(k) = \delta \hat{\rho}_A(k) + \delta \hat{\rho}_B(k)$, and fluctuations in the composition, $\delta\hat{c}(k) = \delta \hat{\rho}_A(k) - \delta \hat{\rho}_B(k)$. These fluctuations can be summarized in terms of the structure factor matrix $\bm{\mathcal{S}}(k)$, defined as
\begin{equation}
\label{eq:structure_factor_matrix}
	\bm{\mathcal{S}}(k) = \begin{pmatrix}
		S_{\rho\rho}(k) & S_{c\rho}(k) \\
		S_{c\rho}(k) & S_{cc}(k)
	\end{pmatrix}
\end{equation}
with matrix elements
\begin{equation}
	\begin{split}
		S_{\rho\rho}(k) &= \frac{1}{V}\,\langle \delta\hat{\rho}(\bm{k}) \, \delta\hat{\rho}(-\bm{k}) \rangle\\
		& = (\rho_{0}^A)^2 \, \hat{h}_{AA}(k) + (\rho_{0}^B)^2 \, \hat{h}_{BB}(k)  \\
		& \quad + \rho_{0}^A + \rho_{0}^B + 2\,\rho_{0}^A\,\rho_{0}^B\,\hat{h}_{AB}(k),
	\end{split}
\end{equation}
\begin{equation}
	\begin{split}
		S_{cc}(k) &= \frac{1}{V}\,\langle \delta\hat{c}(\bm{k}) \, \delta\hat{c}(-\bm{k}) \rangle\\
		& = (\rho_{0}^A)^2 \, \hat{h}_{AA}(k) + (\rho_{0}^B)^2 \, \hat{h}_{BB}(k)  \\
		& \quad + \rho_{0}^A + \rho_{0}^B - 2\,\rho_{0}^A\,\rho_{0}^B\,\hat{h}_{AB}(k),
	\end{split}
\end{equation}
and
\begin{equation}
	\begin{split}
		S_{c\rho}(k) &= S_{\rho c}(k) = \frac{1}{V}\, \langle \delta\hat{c}(\bm{k}) \, \delta\hat{\rho}(-\bm{k}) \rangle\\
		& = (\rho_{0}^A)^2 \, \hat{h}_{AA}(k) - (\rho_{0}^B)^2 \, \hat{h}_{BB}(k)  \\
		& \quad + \rho_{0}^A - \rho_{0}^B .
	\end{split}
\end{equation}
As in equilibrium, we assume that an instability related to a phase transition is signaled by the divergence of fluctuations. To analyze such effects, we diagonalize $\bm{\mathcal{S}}(k)$ and inspect its eigenvalues (or their inverse) and eigenvectors. Symmetric clustering (condensation) is characterized by diverging fluctuations in the total density. A demixing phase transition is signaled by diverging fluctuations in the composition. Consequently, the eigenvalues $\lambda_{1/2}(k)$ and corresponding normalized eigenvectors $\bm{v}_{1/2}(k)=(\delta \hat{\rho}(k),\, \delta \hat{c}(k))^{\rm T}$ of matrix $\bm{\mathcal{S}}(k)$ indicate whether and what type of phase transition is present. More specifically, when $\lambda_1^{-1}(k)$ or $\lambda_2^{-1}(k)$ goes to zero, the respective eigenvector $\bm{v}_{\rm max}$ signals symmetric clustering if $\bm{v}_{\rm max}\approx \bm{x}_{\rm con} = (1,0)^{\rm T}$ or de-mixing if $\bm{v}_{\rm max}\approx \bm{x}_{\rm dem} =(0,1)^{\rm T}$. We quantify the degree of clustering in terms of the angle $\alpha = \arccos(\bm{v}_{\rm max} \cdot \bm{x}_{\rm con})$ between the eigenvector $\bm{v}_{\rm max}$ and the vector $\bm{x}_{\rm con}$, representing symmetric clustering.

Consequently, symmetric clustering is indicated by $\alpha=0$ and symmetric demixing by $\alpha=\pi/2$. Further, the angle $\alpha$ also indicates whether predominantly particles of species $A$ or $B$ form clusters. In particular, $0<\alpha<\pi/2$ corresponds to asymmetric clustering of species $A$, while $-\pi/2 < \alpha < 0$ corresponds to asymmetric clustering of species $B$.

In our analysis, it turns out that the limit $k\rightarrow 0$ is the most relevant since $\lambda_{1/2}^{-1}$ are smallest there. Therefore, the presented results refer exclusively to this limit.

In our calculations, we use the structure factor analysis (and the underlying correlation functions) primarily to characterize the short time behavior after initialization. Specifically, we consider time-averages between $0.5$ and $1\,\tau$ and between $4.5$ and $5\,\tau$.

\section{\label{sec:phase_behavior_stability_diagram}Non-equilibirum phase behavior and stability diagram}
In the present study, we focus on the behavior of ``weakly coupled'' systems, where ``weak'' refers to the values of intraspecies alignment $g_{aa}$. These values are chosen such that the dynamics is \textit{not} dominated by flocking in the entire $g_{AB}-g_{BA}$-plane. Specifically, we set $g_{AA}=g_{BB}=g=3$. This allows us to study the interplay of density and polarization dynamics (rather than the latter alone).

We investigate the phase behavior by performing numerical particle simulations and employing linear stability analyses based on the continuum model at various interspecies coupling strengths $g_{AB}$ and $g_{BA}$. Details regarding the linear stability analysis are presented in Appendix \ref{app:linear_stability_analysis}. It turns out that the linear stability analysis yields accurate predictions in comparison with full non-linear simulations of the continuum equations \cite{kreienkamp_klapp_2022_clustering_flocking_chiral_active_particles_non-reciprocal_couplings}, as shown in Appendix \ref{sapp:continuum_simulations}. Our findings are summarized in the non-equilibrium phase diagram presented in Fig.~\ref{fig:stability_diagram_with_snapshots}.

\subsection{Reciprocal system}
\begin{figure}
	\includegraphics[width=1\linewidth]{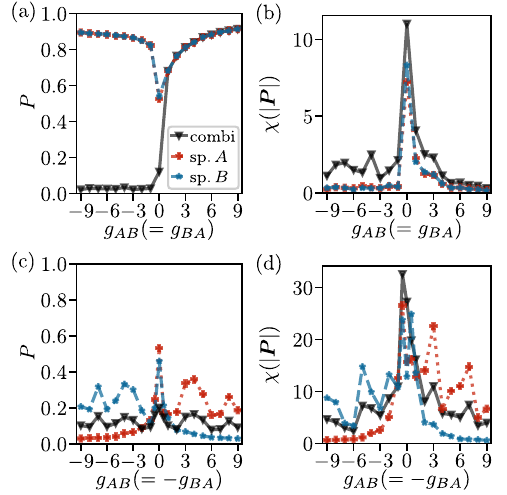}
	\caption{\label{fig:quantities_polarization_particle_simulations} Polarizations $P$ and susceptibilities $\chi(P)$ for (a,b) $g_{AB}=g_{BA}$ (reciprocal case) and (c,d) $g_{AB}=-g_{BA}$ (non-reciprocal case). Data represents time averages of particle simulation results between $70$ and $120\,\tau$ after initialization.}
\end{figure}

\begin{figure}
	\includegraphics[width=1\linewidth]{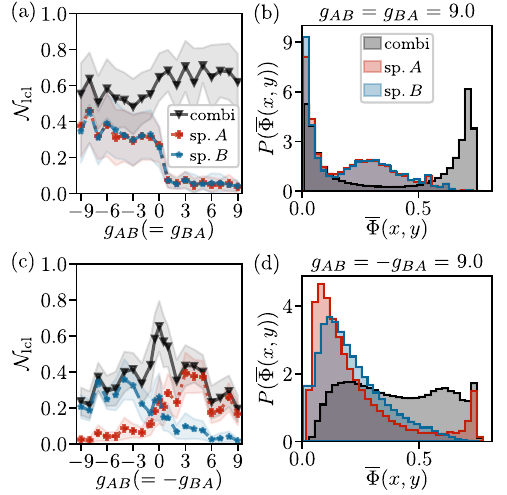}
	\caption{\label{fig:quantities_cluster_formation_particle_simulations} Largest cluster sizes and distributions of local area fraction for reciprocal and non-reciprocal systems. Largest cluster sizes are shown for (a) $g_{AB}=g_{BA}$ and (c) $g_{AB}=-g_{BA}$, calculated as time averages of particle simulation results between $70$ and $120\,\tau$ after initialization. The shaded areas depict the standard deviation of largest cluster sizes within these simulation times. The distributions of the position-resolved local area fraction for (b) $g_{AB}=g_{BA}=9.0$ and (d) $g_{AB}=-g_{BA}=9.0$ represent the time average between $98$ and $100\,\tau$ after initialization.}
\end{figure}

We start by considering the reciprocal system defined by $g_{AB}=g_{BA}=\kappa$.
When couplings between all species are the same, i.e., $g_{ab}=g=3$ ($a,b=\{A,B\}$), the system reduces to an effective one-species system. Choosing this as a starting point [white cross in Fig.~\ref{fig:stability_diagram_with_snapshots}(e)], we now vary $\kappa$, moving along the diagonal white line in Fig.~\ref{fig:stability_diagram_with_snapshots}(e). From particle simulations, we determine the evolution of the polar order parameters, susceptibilities, and largest cluster sizes as a function of the coupling strength $\kappa$. The data are shown in Figs.~\ref{fig:quantities_polarization_particle_simulations}(a,b) and \ref{fig:quantities_cluster_formation_particle_simulations}(a).

\textit{Flocking:}
For $\kappa \gtrsim g$, the system exhibits flocking due to strong alignment. The particles form a large mixed-species cluster consisting of coherently moving particles, as shown in snapshot in Fig.~\ref{fig:stability_diagram_with_snapshots}(a). The flocking behavior is indicated by the large polarization order parameters $P_{\rm combi} = P_A = P_B \gtrsim 0.8$, see Fig.~\ref{fig:quantities_polarization_particle_simulations}(a). The coherent motion of almost all particles in the system yields very small susceptibilities $\chi(P)$ within the flocking regime [Fig.~\ref{fig:quantities_polarization_particle_simulations}(b)].
The cluster formation is reflected by the large percentage of particles in the largest cluster [Fig.~\ref{fig:quantities_cluster_formation_particle_simulations}(a)]. Note that, while the largest cluster size involving all species, $\mathcal{N}_{\rm lcl,\, combi}$, is very large, the corresponding species-resolved values are very small. This expresses the mixed-species character of the cluster. The distributions of local area fraction are shown in Fig.~\ref{fig:quantities_cluster_formation_particle_simulations}(b) and support this observation. The double-peak structure of the distribution of combined local area fraction of all particles clearly indicates the coexistence of a clustered high-density region ($\overline{\Phi}_{\rm combi,cl.}\approx 0.7$) and a dilute low-density region ($\overline{\Phi}_{\rm dilute}\approx 0$). Considering only $A$- or $B$-particles, the distribution still has two peaks but the second (wider) peak has its maximum at $\overline{\Phi}_{A/B,{\rm cl.}}\approx 0.3 \ll \overline{\Phi}_{\rm combi,cl.}$. 

\textit{Phase separation:}
Reducing $\kappa$ such that $\vert \kappa \vert < g$, we enter a regime of weak alignment ($\kappa>0$) or weak anti-alignment ($\kappa<0$) between the two species. The global polar order vanishes since, overall, the alignment couplings become too weak to overcome rotational noise. The system now exhibits pure phase separation, as shown in the snapshot in Fig.~\ref{fig:stability_diagram_with_snapshots}(b).
The decrease of polarization $P_{\rm combi}$ as compared to the flocking regime [Fig.~\ref{fig:quantities_polarization_particle_simulations}(a)] is accompanied by a peak in the susceptibility $\chi(P)$ of the polar order parameter [Fig.~\ref{fig:quantities_polarization_particle_simulations}(b)], indicating a flocking transition at $\kappa\approx0$. At the same time, also the clustering behavior starts to change from mixed-species clusters (where $\mathcal{N}_{{\rm lcl},a}\ll\mathcal{N}_{{\rm lcl, combi}}$) to more demixed, single-species clusters [where $\mathcal{N}_{{\rm lcl},a}\lesssim\mathcal{N}_{{\rm lcl, combi}}$, Fig.~\ref{fig:quantities_cluster_formation_particle_simulations}(a)].
Notably, the particle simulations reveal only a single peak of $\chi(P)$ at $\kappa \approx 0$, suggesting that there is only one orientational transition between the flocking and the anti-flocking regime. This does not quite match the mean-field continuum theory, predicting two flocking transitions, separated by an unpolarized regime, namely, the anti-flocking transition at $\kappa \approx -3$ and the flocking transition at $\kappa \approx 3$. We note in this context that there are other examples in the literature \cite{Farrell_2012_Pattern_formation_self-propelled_particles_density-dependent_motility,liebchen_2017_collective_behavior_chiral_active_matter_pattern_formation_flocking}, where flocking transitions in particle simulations take place at lower coupling strengths than those predicted in the associated continuum models. One reason could be an underestimation of the ``effective'' couplings in the continuum model when considering a small interaction radius.

\textit{Anti-flocking:}
Further reduction of $\kappa$ leads to strong interspecies anti-alignment with $\kappa\lesssim-g$. Here, particles of the same species each form flocks, while flocks of different species move in opposite directions. The anti-parallel orientated flocks consist of single-species clusters, resembling a demixed state, as shown in the snapshot in Fig.~\ref{fig:stability_diagram_with_snapshots}(d). Accordingly, the polarization $P_a$ of either species ({$a=A,B$}) approaches unity, while the polarization $P_{\rm combi}$ of the entire system approaches zero [Fig.~\ref{fig:quantities_polarization_particle_simulations}(a)]. The demixing of particle species is reflected by the largest cluster size with $\mathcal{N}_{{\rm lcl},a}$ approaching $\mathcal{N}_{{\rm lcl, combi}}$ [Fig.~\ref{fig:quantities_cluster_formation_particle_simulations}(a)]. Notably, the standard deviation of the largest cluster size within the anti-flocking regime is relatively large compared to the largest cluster size itself. Accordingly, visual inspection of simulation videos (see movies in \cite{kreienkamp_klapp_PRL}) suggests that single-species clusters in the anti-flocking regime are not very robust, but constantly break up and re-form. This has implications on the global polarization in the system: since a large percentage of particles is quite often {\it not}  part of the largest cluster, ordered regions cannot build up as successfully as in the flocking regime. Therefore, fluctuations of the polarization order parameter are larger. As a result, the susceptibility $\chi(P)$ in the anti-flocking regime is somewhat larger than in the flocking regime [Fig.~\ref{fig:quantities_polarization_particle_simulations}(b)].

\subsection{Non-reciprocal system}
We now explore what happens when we deviate from the reciprocal line, that is, $g_{AB}$ and $g_{BA}$ are considered as mutually independent parameters. 

\begin{figure}
	\includegraphics[width=\linewidth]{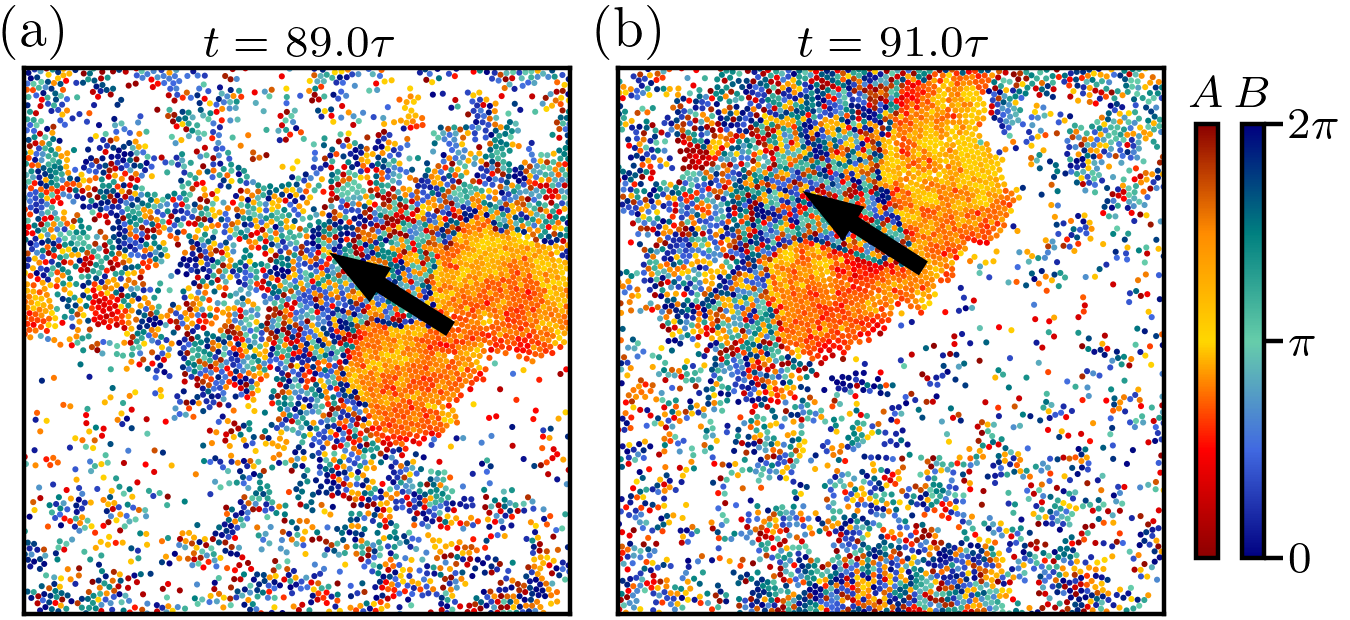}
	\caption{\label{fig:snapshots_chase_and_run} Snapshots illustrating the chase-and-run behavior of asymmetric clusters for $g_{AB}=6$, $g_{BA}=-9$ at two different times (a) $t=89\,\tau$ and (b) $t=91\,\tau$. The black arrow indicates the average orientation within the largest $A$-cluster. Other parameters are $g_{AA}=g_{BB}=3$, ${\rm Pe}=40$, $\Phi=0.4$.}
\end{figure}

Of particular interest are non-reciprocal situations where particles of different species have competing goals. These so-called antagonistic couplings are characterized by different signs of $g_{AB}$ and $g_{BA}$, such that \mbox{$g_{AB}\,g_{BA}<0$}. We focus on the case $g_{AB}=\delta=-g_{BA}$.
From the continuum description, we determine the linear stability of the disordered state, summarized in the non-equilibrium phase diagram in Fig.~\ref{fig:stability_diagram_with_snapshots}(e). From particle simulations, we determine the evolution of the polar order parameters, susceptibilities, and largest cluster sizes as a function of the coupling strength $\delta$. The results are shown in Figs.~\ref{fig:quantities_polarization_particle_simulations}(c,d) and \ref{fig:quantities_cluster_formation_particle_simulations}(c).

\textit{Asymmetric clustering:}
For $\vert\delta\vert \lesssim 20$, the symmetric phase separation of the reciprocal system transforms into an \textit{a}symmetric clustering state, consisting of clusters of only a single species. The characteristic feature of this state is that, depending on the sign of $\delta$, only clusters of species $A$ [snapshot in Fig.~\ref{fig:stability_diagram_with_snapshots}(g)] or $B$ [snapshot in Fig.~\ref{fig:stability_diagram_with_snapshots}(f)] form, independent of initial conditions. This phenomenon is predicted by both, particle simulations and linear stability analyses. Intriguingly, the asymmetrical \textit{density} dynamics arises from non-reciprocal \textit{orientational} couplings between the species whereas all steric interactions are still fully symmetric \cite{kreienkamp_klapp_PRL}. Also, we recall that in our model, orientational couplings between particles are isotropic, i.e., they only depend on the distance, not on the spatial configuration. The microscopic origin of the asymmetric clustering is elucidated in \cite{kreienkamp_klapp_PRL}. The stabilization of clusters of the more aligning species and the dissolution of clusters of the other species is shortly illustrated in Appendix \ref{app:asymmetric_clustering_explanation}. The asymmetric clustering is also reflected by the largest single-species cluster sizes with $N_{{\rm lcl},A}\gg(\ll)N_{{\rm lcl},B}$ for $\delta>(<)0$ [Fig.~\ref{fig:quantities_cluster_formation_particle_simulations}(c)]. The corresponding distributions of local area fractions are shown in Fig.~\ref{fig:quantities_cluster_formation_particle_simulations}(d) for $\delta=9$. The combined-species distribution of local area fraction has no distinct double-peak structure (as in the reciprocal case), but is almost flat. This indicates a coexistence of clustered and loosely accumulated particles. The single-species distributions indeed reflect the partial demixing (i.e., cluster formation) of $A$-particles. These have a typical MIPS-like double-peak structure in $P(\overline{\Phi})$. At the same time, most $B$-particles are in a less dense configuration and only have single-peak in $P(\overline{\Phi})$.
Importantly, while the linear stability analysis can predict the emergence of asymmetric clustering, the \textit{dynamics} of it can only be studied on the particle level. Indeed, particle simulations at these parameters reveal ``run-and-catch'' scenarios, otherwise known from scalar non-reciprocal systems \cite{marchetti_simha_2013_hydrodynamics_soft_active_matter,saha_scalar_active_mixtures_2020,mandal_sollich_2022_robustness_travelling_states_generic_non-reciprocal_mixture,agudo-canalejo_golestanian_2019_active_phase_separation_chemically_interacting_particles,chiu_Omar_2023_phase_coexistence_implications_violating_Newtons_third_law}. The single-species clusters are strongly polarized with polarization $\gtrsim 0.7$. 
The polarized single-species clusters ``chase'' less dense mixed-species accumulations, as shown in snapshots in Fig.~\ref{fig:snapshots_chase_and_run}. The single-species cluster polarization is reflected in the global polarization $P_{A(B)}>P_{B(A)}$ for $\delta >(<) 0$ [Fig.~\ref{fig:quantities_polarization_particle_simulations}(c)]. 
The susceptibilities $\chi(P)$ are shown in Fig.~\ref{fig:quantities_polarization_particle_simulations}(d). They peak at $\delta\approx 0$, which commonly indicates a non-equilibrium phase transition. Within the $A(B)$-clustering regime, the single-species susceptibilities $\chi_{A(B)}(P_{A(B)})$ are relatively large. This reflects strong fluctuations of polarization in the asymmetric clustering regime, which might arise from the fact that not all $a$-particles are part of the asymmetric $a$-clusters. Additionally, simulation videos show that the asymmetric clusters sometimes break up to reassemble quickly into a new cluster of similar size.
We note that the polarization of the single-species clusters is unexpected from the continuum analysis, which only predicts the asymmetric clustering itself. Therefore, also the peaks in the susceptibilities $\chi(P)$ at $\delta \approx 0$ [Fig.~\ref{fig:quantities_polarization_particle_simulations}(b)] are only observed in the particle level approach. 

\begin{figure}
	\includegraphics[width=1\linewidth]{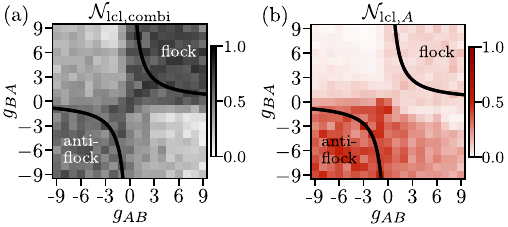}
	\caption{\label{fig:largest_cluster_size_asymmetry}The (scaled) largest cluster size considering (a) all particles, (b) only species $A$. Black lines indicate flocking transition lines from mean-field continuum predictions.}
\end{figure}

\textit{Parameter-dependency of asymmetric clusters:}
When we deviate from the fully anti-symmetric case ($g_{AB}=-g_{BA}$), the linear stability analysis still predicts asymmetric clustering for a broad range of intermediate antagonistic couplings, see Fig.~\ref{fig:stability_diagram_with_snapshots}. Moreover, this method also predicts the \textit{degree} of asymmetric clustering, quantified by the angle $\alpha$ (solid line in Fig.~\ref{fig:stability_diagram_with_snapshots}(c), Appendix \ref{app:linear_stability_analysis}). $A$($B$)-clustering is the most pronounced for $\alpha=(-)45^{\circ}$. Yet, the degree of $A$($B$)-clustering is not uniform but depends on $g_{AB}$ and $g_{BA}$. In particular, towards the flocking transition line, the degree of single-species clustering decreases. This means, for example, that $A$-clustering is predicted to be more pronounced for $g_{AB}=6$, $g_{BA}=-9$ [$\alpha = 43^{\circ}$, snapshot in Fig.~\ref{fig:stability_diagram_with_snapshots}(h)] than for $g_{AB}=9$, $g_{BA}=-6$ [$\alpha = 39^{\circ}$, snapshot in Fig.~\ref{fig:stability_diagram_with_snapshots}(g)]. From particle simulations, we calculate the largest cluster sizes $\mathcal{N}_{\rm lcl, combi}$ and $\mathcal{N}_{{\rm lcl},A}$ for the whole $g_{AB}-g_{BA}$-plane. The results are shown in Fig.~\ref{fig:largest_cluster_size_asymmetry}. We find that $\mathcal{N}_{\rm lcl}$ reflects the trends predicted by the linear stability analysis: within the asymmetrical clustering regimes, $\mathcal{N}_{\rm lcl, combi}$ is larger close to the anti-flocking than flocking regime [Fig.~\ref{fig:largest_cluster_size_asymmetry}(a)]. This can be explained through microscopic considerations outlined in Appendix \ref{app:asymmetric_clustering_explanation}. The comparison with $\mathcal{N}_{{\rm lcl},A}$ [Fig.~\ref{fig:largest_cluster_size_asymmetry}(b)] confirms previous observations: the system exhibits mixed-species clustering in the flocking regime, demixing in the anti-flocking regime, and asymmetrical clustering (whose degree depends on the specific values of $g_{AB},g_{BA}$) in the antagonistic coupling regime. 

\textit{Disorder:}
Finally, for very strong, fully antisymmetric couplings ($\vert\delta\vert \gtrsim 20$), the clustering behavior is suppressed and the homogeneous disordered state emerges [snapshot in Fig.~\ref{fig:stability_diagram_with_snapshots}(i)]. 

\section{\label{sec:correlation_functions}Translational and orientational correlations}

\begin{figure}
	\includegraphics[width=1\linewidth]{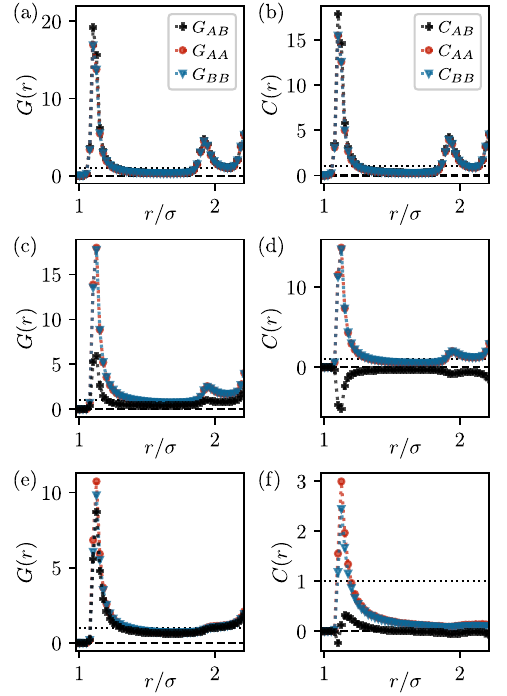}
	\caption{\label{fig:correlation} Positional and orientational pair correlation functions for (a,b) $g_{AB}=g_{BA}=9$, \mbox{(c,d) $g_{AB}=g_{BA}=-9$,} and \mbox{(e,f) $g_{AB}=-g_{BA}=9$.} The intraspecies alignment strength is $g_{AA}=g_{BB}=3$. The correlation functions represent time averages between $0.5$ and $1\,\tau$ after initialization.} %
\end{figure}

To further explore the microscopic structure we now discuss positional and orientational pair correlation functions as defined in Eqs.~\eqref{eq:pair_distribution_function} and \eqref{eq:orientational_correlation}. We focus on the emergence of correlations after initialization from a disordered state and, thus, on short times.
For simplicity, we here neglect the angle-dependency of the pair correlation function and consider directly the angle-averages $G_{ab}(r)= \langle G_{ab}(r,\phi) \rangle_{\phi}$ and $C_{ab}(r)= \langle C_{ab}(r,\phi) \rangle_{\phi}$. Results are shown in Fig.~\ref{fig:correlation}.

In case of strong {\it reciprocal} alignment ($g=3$, $\kappa=9$), the system exhibits a phase-separated flocking state (see Sec.~\ref{sec:phase_behavior_stability_diagram}). Here, the short-time pair correlation functions $G_{ab}(r)$ have pronounced first peaks at about one particle diameter, followed by smaller double peaks around $r/\sigma \approx 2$ [Fig.~\ref{fig:correlation}(a)]. The latter feature is characteristic of a clustering of particles in locally hexagonal arrangements. At large distances, the $G_{ab}(r)$ tends to unity, as expected. 
Noticeably, in this reciprocal case with $\kappa > g$, the two intraspecies correlations are the same, $G_{AA}=G_{BB}$, while the interspecies correlations differ, $G_{AB} > G_{AA}$. The reason is that alignment promotes clustering \cite{elena_2018_velocity_alignment_MIPS}, such that species $a$ is more likely to be surrounded by species $b\neq a$ than by the same species. This reflects the mixed-species clustering. For the depicted small particle distances, the corresponding orientational correlation functions $C_{ab}(r)$ show very similar features as $G_{ab}(r)$ [Fig.~\ref{fig:correlation}(b)], indicating the parallel orientation of close-by particles in the flocking case. At large distances, $C_{ab}(r)$ tends to zero (not shown). This is due to the short times considered, where orientational correlations do not yet span over the entire system.

We now consider strong reciprocal anti-alignment ($g=3$, $\kappa=-9$), where the system exhibits a demixed anti-flocking state (see Sec.~\ref{sec:phase_behavior_stability_diagram}). The pronounced peaks in $G_{ab}(r)$ indicate again a clustering of particles [Fig.~\ref{fig:correlation}(c)]. However, contrary to the flocking case, the anti-alignment between particles of different species results in $G_{AB}<G_{AA}=G_{BB}$, reflecting the demixing. The corresponding orientational correlation functions [Fig.~\ref{fig:correlation}(d)] support the observation that particles of the same species are aligned ($C_{aa}\approx G_{aa}$, $a=A,B$), whereas particles of different species are anti-aligned ($C_{AB}\approx -G_{AB}$). At large distances, the short-time orientational correlations vanish.

Finally, in case of {\it non-reciprocal} (anti-)alignment ($\delta=9$), the pair correlation functions exhibit a distinctive feature: the intraspecies correlations are no longer the same, i.e., $G_{AA}\neq G_{BB}$ [Fig.~\ref{fig:correlation}(e)]. Further, while $G_{ab}(r)$ still exhibits a pronounced first peak, the peaks at larger distances disappear. This indicates that the system is, overall, in a more disordered state compared to the cases discussed before. The asymmetry $G_{AA}> G_{BB}$ in the first peak is related to the asymmetric cluster growth observed in longer particle simulations. Indeed, for the present parameters, particles of species $A$ start clustering, while $B$-particles only form loose accumulations. Finally, the orientational correlations [Fig.~\ref{fig:correlation}(f)] express the enhanced alignment of species $A$ as compared to $B$ ($C_{AA}>C_{BB}$), while there are essentially no correlations between particles of different species ($C_{AB}\approx 0$).

To summarize this paragraph, we have seen that pair correlation functions, that are calculated shortly after the initialization of the simulation, can successfully predict the emerging clustering and local (anti-)alignment. In particular, these functions reveal the asymmetry of dynamical structures when non-reciprocal interactions are at play. This type of information is clearly not available in the simpler mean-field continuum approach.

\section{\label{sec:structure_factors}Structure factor analysis}
\begin{figure}
	\includegraphics[width=1\linewidth]{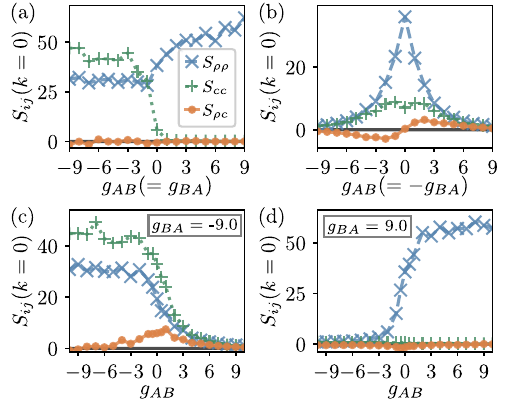}
	\caption{\label{fig:structure_factor_matrix_elements}Elements of the structure factor matrix [Eq.~\eqref{eq:structure_factor_matrix}] for (a) $g_{AB}=g_{BA}$, (b) $g_{AB}=-g_{BA}$, (c) $g_{AB}=-9$, and (d) $g_{AB}=9$. The intraspecies coupling is $g_{AA}=g_{BB}=3$. The matrix elements are calculated as a time average between $4.5$ and $5\,\tau$ after initialization.}
\end{figure}

\begin{figure}
	\includegraphics[width=1\linewidth]{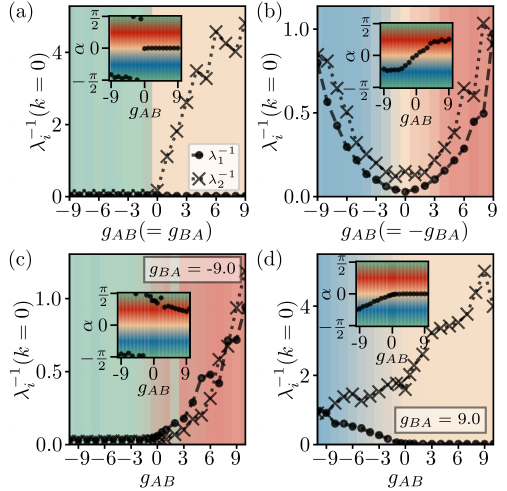}
	\caption{\label{fig:structure_factor_matrix_eigenvalues}Eigenvalues of the structure factor matrix (Eq.~\eqref{eq:structure_factor_matrix}) for (a) $g_{AB}=g_{BA}=\kappa$, (b) $g_{AB}=-g_{BA}=\delta$, (c) $g_{AB}=-9$, and (d) $g_{AB}=9$. The intraspecies coupling is $g_{AA}=g_{BB}=3$. The matrix elements are calculated as a time average between $4.5$ and $5\,\tau$ after initialization.}
\end{figure}

Given the detailed structural description provided by the correlation functions (Sec.~\ref{sec:correlation_functions}), we now use these quantities as an input for a systematic fluctuation analysis. 
This yields further quantitative information regarding the phase separation behavior of the binary mixture. 
To this end, we calculate the two-dimensional matrix $\bm{\mathcal{S}}(k)$ of structure factors, i.e., density fluctuations, at $k=0$ as outlined in Sec.~\ref{ssec:analysis_methods_structure_factor_matrix}. The calculations of $\bm{\mathcal{S}}(k)$ are done at short times. Examples for the elements of $\bm{\mathcal{S}}(k)$ as a function of the orientational coupling parameters are shown in Fig.~\ref{fig:structure_factor_matrix_elements}. We then extract the (inverse) eigenvalues $\lambda_{1/2}^{-1}$ and related angles $\alpha$ from the eigenvectors to obtain a prediction for the type of phase separation. Data are shown in Fig.~\ref{fig:structure_factor_matrix_eigenvalues}.

We first consider the reciprocal line with $g_{AB}=g_{BA}=\kappa$. Depending on $\kappa$, the mean-field linear stability analysis predicts anti-flocking, phase separation without flocking, and flocking (Sec.~\ref{sec:phase_behavior_stability_diagram}). The fluctuation analysis reveals that within the anti-flocking regime ($\kappa<0$), composition fluctuations (i.e., $\mathcal{S}_{cc}$) dominate, whereas in the flocking regime ($\kappa>0$), density fluctuations (i.e., $\mathcal{S}_{\rho\rho}$) become prominent [Fig.~\ref{fig:structure_factor_matrix_elements}(a)]. Mixed density-concentration fluctuations $\mathcal{S}_{\rho c}$ are around zero for all $\kappa$. Consistent with these finding, the eigenvalue and eigenvector analysis predicts an instability ($\lambda_1^{-1}\approx 0$) and the transition from a demixing ($\alpha \approx \pm \pi/2$) towards a symmetric clustering instability ($\alpha \approx 0$) at $\kappa \approx 0$ [Fig.~\ref{fig:structure_factor_matrix_eigenvalues}(a)].

We now consider the case of fully anti-symmetric non-reciprocal couplings, i.e., $g_{AB}=-g_{BA}=\delta$. Here, we find $\mathcal{S}_{\rho\rho}>\mathcal{S}_{cc}$ for all $\delta$ [Fig.~\ref{fig:structure_factor_matrix_elements}(b)]. Furthermore, mixed density-composition fluctuations $\mathcal{S}_{\rho c}$ are non-zero and change sign at the same time as $\delta$ changes sign. The inverse eigenvalues in the non-reciprocal case are non-zero for all finite values of $\delta$ [Fig.~\ref{fig:structure_factor_matrix_eigenvalues}(b)], expressing the fact that all density fluctuations are large, but not divergent. The eigenvector angle of $\alpha \approx -(+) \pi/4$ indicates asymmetric $B$-\mbox{($A$-)}clustering for $\delta <(>) 0$, consistent with our previous observations. 

Setting $g_{BA}=-9$ and increasing $g_{AB}$ from the same value, we move along an horizontal line in the stability diagram Fig.~\ref{fig:stability_diagram_with_snapshots}(e) from the anti-flocking to the asymmetric $A$-clustering phase. Along this way, pure composition fluctuations, $\mathcal{S}_{cc}$, dominate [Fig.~\ref{fig:structure_factor_matrix_elements}(c)]. Further, mixed fluctuations, $\mathcal{S}_{\rho c}$, become non-negative for small $\vert g_{AB} \vert$. Correspondingly, the eigenvalues and -vectors indicate a transition from a demixed state (related to anti-flocking), to asymmetric $A$-clustering [Fig.~\ref{fig:structure_factor_matrix_eigenvalues}(c)]. 

As a final example, we consider $g_{BA}=9$. Here, pure density fluctuations are found to dominate for all $g_{AB}$ [Fig.~\ref{fig:structure_factor_matrix_elements}(d)]. The transition from the asymmetric $B$-clustering to a symmetric phase separation seen in Fig.~\ref{fig:stability_diagram_with_snapshots}(e) is reproduced by the eigenvalues and -vectors [Fig.~\ref{fig:structure_factor_matrix_eigenvalues}(d)]. Noticeably, the non-zero (negative) mixed fluctuations, $\mathcal{S}_{\rho c}$, which accompany the asymmetric clustering transition, are significantly less pronounced than in the $g_{BA}=-9$ case [Fig.~\ref{fig:structure_factor_matrix_elements}(c)]. This reflects the dependency of asymmetric clustering (and related cluster sizes, Fig.~\ref{fig:largest_cluster_size_asymmetry}) not only on the sign of $g_{AB}$ and $g_{BA}$, but also on their ratio.

In this section, we have shown that a systematic analysis of long-wavelength fluctuations yields a reliable route towards state transformations in agreement with the preceding particle-level analysis and mean-field results (where available). The fluctuation analysis additionally provides detailed information on the clustering behavior.

In fact, in some cases, we find quantitative agreement when comparing fluctuation and mean-field stability analysis. In Fig.~\ref{fig:stability_diagram_with_snapshots}(c), we have plotted the phase separation angle $\alpha$ as a function of $\delta$. It is seen that both methods consistently predict a gradual transition from asymmetric $B$-clustering to symmetric clustering and, finally, asymmetric $A$-clustering.

This very good agreement is indeed not obvious. Within the linear stability analysis, clustering is detected at finite wavelength ($k>0$), in contrast to polarization instabilities [(anti-)flocking] that occur at $k=0$ \cite{kreienkamp_klapp_2022_clustering_flocking_chiral_active_particles_non-reciprocal_couplings}. In turn, our analysis of density fluctuations, which is based on short-time correlation functions, pertains to $k=0$. By this we monitor long-range correlations of fluctuations rather than instabilities of $k=0$-related mean values in the continuum description.

\section{\label{sec:no_repulsion}Particle-simulations without repulsion}
An important question is to what extend the asymmetric clustering is a result of the mixed occurrence of repulsion and alignment couplings. Indeed, in one-component Vicsek-like systems, (reciprocal) alignment is known to induce inhomogeneous density states -- even in the absence of MIPS. In simple flocking models of point-like particles, regimes of high-density polarized bands separate homogeneous disorder and homogeneous flocking states \cite{vicsek_1995_novel_type_phase_transition,solon_tailleur_2015_phase_separation_flocking_models}. Hence, we now address the question of how non-reciprocal alignment affects the density dynamics in systems without repulsion.

To this end, we perform BD simulations of non-repulsive particles ($\epsilon=0$) in the same ``weakly-coupled'' regime as considered before, i.e., $g=3$. Without repulsion, no self-trapping mechanism can induce MIPS. Then, all accumulations of particles can be attributed to the alignment couplings alone.

\begin{figure}
	\includegraphics[width=1\linewidth]{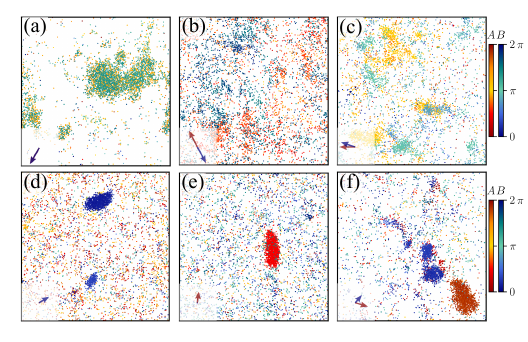}
	\caption{\label{fig:snapshots_no_repulsion}Snapshots of BD simulations without repulsion for (a) $g_{AB}=g_{BA}=9$ (flocking), (b) $g_{AB}=g_{BA}=-9$ (anti-flocking), (c) $g_{AB}=g_{BA}=0$ (non-interacting species), (d) $g_{AB}=-g_{BA}=-9$ ($B$-cluster), (e) $g_{AB}=-g_{BA}=9$ ($A$-cluster), and (f) $g_{AB}=-g_{BA}=3$ ($A$-cluster). Small arrows in lower left corner of snapshots indicate the average polarization of both species.}
\end{figure}

\begin{figure}
	\includegraphics[width=1\linewidth]{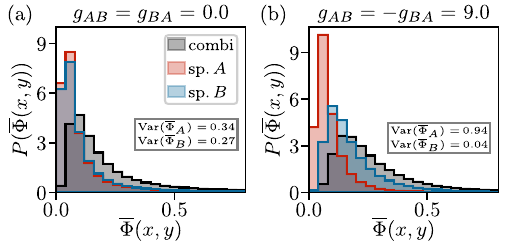}
	\caption{\label{fig:density_distribution_no_repulsion} Distributions of local area fraction for reciprocal and non-reciprocal systems without repulsion for (a) $g_{AB}=g_{BA}=0.0$ and (b) $g_{AB}=-g_{BA}=9.0$. Data represents the time average between $38$ and $40\,\tau$ after initialization.}
\end{figure}

In Fig.~\ref{fig:snapshots_no_repulsion}, BD simulation snapshots of non-repulsive point-particles are shown for different $g_{AB}, g_{BA}$, while diffusion and density are kept as before. Distributions of position-resolved local area fractions are shown in Fig.~\ref{fig:density_distribution_no_repulsion}.

The reciprocal flocking state at $g_{AB}=g_{BA}=9$ [Fig.~\ref{fig:snapshots_no_repulsion}(a)] consists of mixed-species, high-density clumps with large polarization. At $g_{AB}=g_{BA}=-9$ [Fig.~\ref{fig:snapshots_no_repulsion}(b)], anti-flocking with a more homogeneous particle density emerges. Here, both species form flocks, yet with anti-parallel orientations. Generally, anti-flocking states can consist of anti-parallel polarized bands or homogeneous anti-parallel flocks depending on the effective alignment strengths \cite{Chatterjee_Noh_2023_flocking_two_unfriendly_species}. For non-interacting species [Fig.~\ref{fig:snapshots_no_repulsion}(c)] with $g_{AB}=g_{BA}=0$, both species individually form inhomogeneous flocks, whose direction is not correlated. An important difference between the systems with and without repulsion is the emergence of MIPS. Without repulsion, the particles accumulate only due to alignment. The distribution of local area fraction does not have the MIPS-typical two-peak structure [compare Fig.~\ref{fig:quantities_cluster_formation_particle_simulations}(b)], but only features a single peak, see Fig.~\ref{fig:density_distribution_no_repulsion}(a).

Snapshots of non-reciprocal systems are shown in Figs.~\ref{fig:snapshots_no_repulsion}(d)-(f). Depending on the strength of $g_{AB}$ and $g_{BA}$, different asymmetric accumulation behavior is observed. For $g_{AB}=-g_{BA}=\pm 9$ [Figs.~\ref{fig:snapshots_no_repulsion}(d),(e)], only particles of species $A$ or $B$ form polarized clumps. In Fig.~\ref{fig:density_distribution_no_repulsion}(b), the distribution of local area fraction is shown for asymmetric $A$-clustering. The strong accumulation of $A$-particles at single points cannot easily be seen in the distribution but is clearly reflected in the high variance of local area fractions of $A$-particles. This single-species clustering is in accordance with results including repulsion. Yet, without repulsion, there is no chase-and-run behavior (compare Fig.~\ref{fig:snapshots_chase_and_run}).

Another stark difference is observed for weaker non-reciprocity, e.g., $g_{AB}=-g_{BA}=-3$ [Fig.~\ref{fig:snapshots_no_repulsion}(f)]. Here, not only the expected $A$-``clustering'' emerges but also $B$-particles start to accumulate. In systems with repulsion, the clustering of the anti-aligning $B$-species due to orientational couplings is hindered. The reason is that repulsive interactions lead to trapping of close-by particles. These particles then have more time to impact the other particles' orientations, thereby destroying $B$-clusters as described in \cite{kreienkamp_klapp_PRL} and Appendix \ref{app:asymmetric_clustering_explanation}. The trapping mechanism is not present in systems without repulsion. Here, $B$-particles accumulate because of alignment within the species ($g_{BB}>0$) and are not heavily affected by the relatively weak orientational couplings to $A$-particles.

Thus, non-reciprocal polar alignment alone induces asymmetric clustering. However, the overall density dynamics are heavily affected.

\section{\label{sec:summary_conclusion}Summary and conclusions}
In this paper, we investigated dynamical structures and pattern formation in a binary mixture of mutually repelling and non-reciprocally aligning particles using methods on different scales. While our mean-field hydrodynamic description already identifies instabilities and emerging large-scale patterns \cite{kreienkamp_klapp_PRL}, our primary focus in the present paper was to describe and understand the emerging translational and orientational structures from a microscopic particle-based perspective.

To this end, we performed extensive particle simulations based on the Langevin equations of the system. Calculating order parameters, susceptibilities, and pair correlations allowed us to characterize different non-equilibrium phases and associated transitions, which were then compared to the predictions of the continuum model.

As already outlined in \cite{kreienkamp_klapp_PRL}, the \textit{reciprocal} system exhibits symmetric (anti-)flocking and phase separating states, while antagonistic \textit{non-reciprocal} orientational couplings induce asymmetric clustering of only one particle species. This is indeed remarkable given that the translational interactions in our model are fully symmetric (as are the dynamical parameters) and do no couple to the orientations. The here presented particle-based analysis shows that the emerging single-species clusters are strongly polarized, while the full system does not feature long-range order. Further, we observe a ``chase-and-run'' behavior, where the single-species clusters chase the more loosely accumulated particles of the other species. The asymmetry in cluster formation can be explained on the microscopic level by the stabilization of clusters of the stronger aligning species, while clusters of the anti-aligning species dissolve. Interestingly, this phenomenon is already reflected in the pair correlations shortly after the initialization of the system. Consequently, our fluctuation analysis based on density-composition structure factors provides a very accurate prediction of the clustering phenomenon.

From a methodological point of view, the present study confirms that the mean-field continuum theory does not only predict the non-equilibrium phases seen in the Brownian dynamics simulations. In fact, to some extent there is also quantitative agreement, a prime example being the degree of asymmetric clustering [Fig.~\ref{fig:stability_diagram_with_snapshots}(c)]. On the other side, as expected, the particle simulations provide additional structural and dynamical insights that surpass the mean-field continuum description.

Our findings are relevant to various real non-equilibrium systems, where non-reciprocity is, in fact, quite common. Our results could, in principle, be tested in engineered robotic experiments \cite{lavergne_bechinger_2019_group_formation_visual_perception-dependent_motility,wang_cichos_2023_spontaneous_vortex_formation_retared_attractions}, where non-reciprocal alignment rules have already been implemented \cite{chen_zhang_2024_emergent_chirality_hyperuniformity_active_mixture_NR}. Another experimental realization is given by mixture of ``colloidal Quincke rollers'' \cite{maity_morin_2023_spontaneous_demixing_binary_colloidal_flocks}. Quincke rollers are insulating particles, dissolved in a conducting fluid, which start to self-propel when an electric field is applied \cite{bricard_2013_emergence_macroscopic_directed_motion}. Interactions between these particles comprise steric repulsion and \mbox{(anti-)}alignment torques induced by hydrodynamic and electrical couplings. For colloidal rollers of different sizes, it has been shown that the \mbox{(anti-)}alignment torques become non-reciprocal, leading to spontaneous demixing \cite{maity_morin_2023_spontaneous_demixing_binary_colloidal_flocks}. This experiment demonstrates the coupling between (non-reciprocal) orientational torques and demixing. Yet, while Quincke rollers interact via (more complex) hydrodynamic and electrical dipolar interactions, our model comprises much ``simpler'', Vicsek-like torques. Understanding the impact of non-reciprocal orientational couplings on repulsive systems or systems with other isotropic interactions is therefore essential for gaining insights into the overall collective behavior of such systems.

\begin{acknowledgments}
This work was funded by the Deutsche Forschungsgemeinschaft (DFG, German Research
Foundation) -- Projektnummer 163436311 (SFB 910) and Projektnummer 517665044.
\end{acknowledgments}

\appendix

\section{\label{app:continuum_model}Continuum model}
In this Appendix we provide information on the mean-field continuum model with which our particle-based results are compared. The essential steps of the derivation of the continuum model have been described in detail in \cite{kreienkamp_klapp_2022_clustering_flocking_chiral_active_particles_non-reciprocal_couplings}. Here, we give a brief summary.

\subsection{\label{sapp:steric_effects_continuum}Steric effects}
We start with our treatment of steric effects in the continuum theory. Previous numerical \cite{Bialke_2013_microscopic_theory_phase_seperation,elena_2021_phase_separation_self-propelled_disks} and analytical \cite{Bialke_2013_microscopic_theory_phase_seperation,speck_2015_dynamical_mean_field_phase_separation, damme_2019_interparticle_torques_phase_separation,elena_2021_phase_separation_self-propelled_disks} studies have shown that self-propulsion introduces a force imbalance for systems of particles of finite size, since there are more particles in front than behind a reference particle. The force imbalance leads to an effective velocity reduction depending on the density of surrounding particles. However, the mean-field
character of our continuum theory disregards the structure of pair correlations, which become anisotropic due to activity. As a result, on the mean-field continuum level, a constant self-propulsion speed does not reproduce motility-induced phase separation (MIPS) into low- and high-density regions \cite{buttinoni_2013_dynamical_clustering,van_der_linden_2019_interrupted_mips,bauerle_2018_self-organization_quorum_sensing,liu_2019_self-driven_phase_transitions,OByrne_Zhao_2023_introduction_to_MIPS, kreienkamp_klapp_2022_clustering_flocking_chiral_active_particles_non-reciprocal_couplings}. 

In order to account for this characteristic phenomenon in an interacting repulsive system, we employ an \textit{effective} density-dependent velocity of the active particles on the continuum level, rather than the constant speed $v_0$ appearing in LE \eqref{eq:Langevin_eq} \cite{Bialke_2013_microscopic_theory_phase_seperation,speck_2015_dynamical_mean_field_phase_separation,damme_2019_interparticle_torques_phase_separation,cates_tailleur_2013_ABP_run-and-tumble_equivalent,Farrell_2012_Pattern_formation_self-propelled_particles_density-dependent_motility,elena_2021_phase_separation_self-propelled_disks,worlitzer_2021_mips_meso-scale_turbulence_active_fluids}. Specifically, we assume that the effective velocity is given by \cite{Bialke_2013_microscopic_theory_phase_seperation,speck_2015_dynamical_mean_field_phase_separation,damme_2019_interparticle_torques_phase_separation,elena_2021_phase_separation_self-propelled_disks,worlitzer_2021_mips_meso-scale_turbulence_active_fluids}
\begin{equation}
	\label{eq:effective_velocity}
	v = v^{\rm eff}(\rho)=v_0 - \zeta \, \rho .
\end{equation}
Eq.~\eqref{eq:effective_velocity} expresses the fact that particles are slowed down in crowded situations, depending on the overall local particle density, $\rho=\rho(\bm{r})=\sum_a \rho^a$, and the ``friction''-like velocity-reduction parameter, $\zeta$.

Note that the ansatz \eqref{eq:effective_velocity} disregards the effect of alignment couplings on the effective velocity (which are, in fact, non-negligible for $g_{AB},g_{BA}\geq 3$). As indicated by our particle-simulations, the density-dependent decrease is, strictly speaking, only observed for anti-alignment or non-reciprocal couplings. Still, Eq.~\eqref{eq:effective_velocity} provides a convenient starting point for handling repulsive forces. 

The Langevin equations from which the continuum model is derived are therefore given by
\begin{subequations} \label{eq:Langevin_eq_for_continuum}
	\begin{align}
		\dot{\bm{r}}_{\alpha}(t) &= v^{\rm eff}(\rho)\,\bm{p}_{\alpha}+ \bm{\xi}_{\alpha}(t)\\
		\dot{\theta}_{\alpha}(t) &= \mu_{\theta} \sum_{\beta\neq\alpha} \mathcal{T}_{\rm al}^{\alpha}(\bm{r}_{\alpha}, \bm{r}_{\beta}, \theta_{\alpha}, \theta_{\beta}) + \eta_{\alpha}(t)
	\end{align}
\end{subequations}
with the torque
\begin{equation}
	\label{eq:torque_for_continuum}
	\mathcal{T}_{\rm al}^{\alpha}(\bm{r}_{\alpha}, \bm{r}_{\beta}, \theta_{\alpha}, \theta_{\beta}) = k_{ab}\, \sin(\theta_{\beta}-\theta_{\alpha}) \, \Theta(R_{\theta}-r_{\alpha\beta}).
\end{equation}

\cor{Passive particles with vanishing self-propulsion velocity $v_0 = 0$ in LE \eqref{eq:Langevin_eq} are still subject to steric repulsion. However, in LE \eqref{eq:Langevin_eq_for_continuum}, used to derive the associated continuum model, the passive limit translates into $v^{\rm eff}=0$. As a result, particle positions are only subject to translational noise and steric repulsion between particles is neglected in the continuum model in the passive limit. This is motivated by the fact that we are interested in density regimes, where the passive particle system with short-ranged steric repulsion shows no clustering or phase separation, see Supplemental Material of \cite{kreienkamp_klapp_PRL}. Depending on the \mbox{(non-)}reciprocal alignment, the passive system shows qualitative similarities to a classical Heisenberg fluid in a paramagnetic or ferromagnetic fluid phase without gas-liquid phase separation \cite{hoye_stell_1976_configurationally_disordered_spin_systems,oukouiss_baus_1997_phase_diagrams_classical_heisenberg_fluid}. In principle, short-range steric repulsion could be described on a continuum level in terms of free-energy functionals in the framework of classical dynamical density functional theory \cite{marconi_tarazona_1999_dynamic_density_functional_theory_fluids,archer_2004_dynamical_density_function_theory} or density-dependent diffusion coefficients \cite{felderhof_1978_diffusion_interaction_Brownian_particles}. Nevertheless, in the parameter regime of interest, the resulting dynamics would not show any clustering effects in the passive limit, consistent with the associated particle system.}

\subsection{Derivation}
To derive the continuum model, we closely follow the steps presented in \cite{fruchart_2021_non-reciprocal_phase_transitions,kreienkamp_klapp_2022_clustering_flocking_chiral_active_particles_non-reciprocal_couplings,teVrugt_Wittkowski_2023_derivation_predictive_field_theory}. The first step is a mean-field Fokker-Planck equation for the one-particle probability density function (PDF)
\begin{equation}
	f^a(\bm{r},\theta,t) = \frac{1}{N_a} \sum_{\alpha}^{N_a} \langle\delta(\bm{r}-\bm{r}_{\alpha}(t)) \, \delta(\theta - \theta_{\alpha}(t))\rangle.
\end{equation}
The resulting Fokker-Planck equation is given by
\begin{equation}
	\label{eq:Fokker-Planck_equation}
	\begin{split}
		\frac{\partial}{\partial t} f^{a}(\bm{r},\theta,t)
		= & - \nabla \cdot \Big\{f^{a}(\bm{r},\theta,t)\, v^{\rm eff}(\rho)\,\bm{p}(\theta) \Big\} \\
		& - \partial_{\theta} \,\Big\{f^{a}(\bm{r},\theta,t) \, \mu_{\theta} \, \mathcal{I}_f(r,\theta,t) \Big\} \\
		& + \xi \, \nabla^2 \, f^{a}(\bm{r},\theta,t)+ \eta \, \partial_{\theta}^2 \, f^{a}(\bm{r},\theta,t) ,
	\end{split}
\end{equation}
where $\nabla$ and $\partial_{\theta}$ denote derivatives in space ($\bm{r}$) and orientation angle ($\theta$), respectively, and $\mathcal{I}_f(r,\theta,t)$ denotes the integral 
\begin{equation}
	\label{eq:integral_in_FP}
	\begin{split}
		&\mathcal{I}_f(r,\theta,t) = R_{\theta}^2 \int \sum_b N_b\, k_{ab} \, {\rm sin}(\theta'-\theta)\, f^b(\bm{r},\theta',t)\,{\rm d}\theta'.
	\end{split}
\end{equation}
Here, we used the mean-field approximation, which assumes that the two-particle PDF $f(\bm{r},\bm{r}',\theta,\theta',t)$ is the product of two one-particle PDFs. We further replaced the step function $\Theta(r_{\theta}-r_{\alpha\beta})$ by $R_{\theta}^2\,\delta(\bm{r}_{\alpha}-\bm{r}_{\beta})$. This is justified if the interaction radius is small enough \cite{fruchart_2021_non-reciprocal_phase_transitions,yllanes_2017_dissenters_to_disorder_a_flock,liebchen_2017_collective_behavior_chiral_active_matter_pattern_formation_flocking,mietke_dunkel_2022_anyonic_defect_braiding}.

To evaluate the remaining orientational integral in Eq.~\eqref{eq:integral_in_FP} and derive time-evolution equations for (orientational) moments, we follow the approach outlined in \cite{Bertin_2009_hydrodynamic_equations_self-propelled_particles} and express the one-particle PDF in terms of its Fourier expansion with respect to the angle $\theta$, i.e.,
\begin{equation}
	f^a(\bm{r},\theta,t) = \frac{1}{2\,\pi} \, \sum_{n=-\infty}^{\infty} \hat{f}_n^a(\bm{r},t) \, {\rm e}^{-in\theta}.
\end{equation}
The time-evolution of coefficients is given by \cite{kreienkamp_klapp_2022_clustering_flocking_chiral_active_particles_non-reciprocal_couplings}
\begin{equation}
	\label{eq:Fourier_modes_time_evolution}
	\begin{split}
		\partial_t & \hat{f}_n^a \\
		= & -\frac{1}{2} \left[ \partial_z \, (v^{\rm eff}(\rho)\,\hat{f}_{n-1}^a) +  \partial_{\overline{z}}\,(v^{\rm eff}(\rho)\, \hat{f}_{n+1}^a) \right] \\
		& +\frac{R_{\theta}^2\,\mu_{\theta}}{2}  \sum_b N_b \, 	k_{ab} \, n\,  \Big\{ \hat{f}_{n-1}^a\, \hat{f}_{1}^b - \hat{f}_{n+1}^a \, \hat{f}_{-1}^b \Big\}\\
		& - \xi \,  \partial_z\,\partial_{\overline{z}} \, \hat{f}_n^a - \eta \,  n^2 \, \hat{f}_n^a,
	\end{split}
\end{equation}
where $\partial_z = \partial_x + i\,\partial_y$ and $\partial_{\overline{z}} = \partial_x - i \, \partial_y$ \cite{fruchart_2021_non-reciprocal_phase_transitions}.
The Fourier modes can be related to moments of the one-particle PDF $f^a(\bm{r},\theta,t)$. In particular, the particle density (related to mode $n=0$) is given by
\begin{equation}
	\label{eq:density_field_by_f0}
	\begin{split}
		\rho^a(\bm{r},t) &= N_a \, \hat{f}^a_0(\bm{r},t) = N_a \int_{-\pi}^{\pi} f^a(\bm{r},\theta,t)\,{\rm d}\theta.
	\end{split}
\end{equation}
Next, the polarization density (related to mode $n=1$) is defined as
\begin{equation}
	\begin{split}
		\bm{w}^a(\bm{r},t) &= N_a  \begin{pmatrix}
			{\rm Re}(\hat{f}_1^a) \\ {\rm Im}(\hat{f}^a_1)
		\end{pmatrix} \\
		&= N_a \int_{-\pi}^{\pi} f^a(\bm{r},\theta,t) \,\bm{p}(\theta) \, {\rm d}\theta,
	\end{split}
\end{equation}
describing the average orientation of particles of species $a$ via $\bm{w}^a/\rho^a$.

The time evolution \eqref{eq:Fourier_modes_time_evolution} of the Fourier modes $\hat{f}_n^a$ represents a hierarchical set of equations, necessitating the use of a consistent closure scheme.
Here, we use a scaling ansatz \cite{Bertin_2009_hydrodynamic_equations_self-propelled_particles, Bertin_2006_Boltzmann_hydrodynamic_self-propelled_particles}, yielding
\begin{equation}
	\label{eq:approximation_moment_f2}
	\begin{split}
		\hat{f}_2^a = -\frac{1}{4\,\eta} \Bigg (&\frac{1}{2}\, \partial_z \,(v^{\rm eff}(\rho)\,\hat{f}_1^a) \\
		&- R_{\theta}^2\, \mu_{\theta} \, \sum_b N_b \, k_{ab} \, \hat{f}_1^a \, \hat{f}_1^b \Bigg ).
	\end{split}
\end{equation}
By applying the closure relation, the full dynamics of the one-particle PDF is reduced to the dynamics of the particle and polarization density. 

\subsection{\label{sapp:full_continuum_equations}Continuum equations}
Using Eqs.~\eqref{eq:Fourier_modes_time_evolution} and \eqref{eq:approximation_moment_f2}, the evolution equation for the density fields \eqref{eq:density_field_by_f0} becomes
\begin{equation}
	\label{eq:continuum_eq_density}
	\partial_t \rho^a +  \nabla \cdot \bm{j}_{a} = 0
\end{equation}
with flux
\begin{equation}
	\label{eq:continuity_flux}
	\bm{j}_a = v^{\rm eff}(\rho)\, \bm{w}^a -  D_{\rm t}\,\nabla\,\rho^a .
\end{equation} 
The flux involves the polarization density $\bm{w}^a$, which evolves according to
{\small 
	\begin{equation}
		\label{eq:continuum_eq_polarization}
		\begin{split}
			\partial_t & \bm{w}^a  \\
			=& - \frac{1}{2} \, \nabla \,\big(v^{\rm eff}(\rho)\, \rho^a\big)  - D_{\rm r }\, \bm{w}^{a} + \sum_b g'_{ab} \, \rho^a\, \bm{w}^b \\
			&+  D_{\rm t}\,\nabla^2\,\bm{w}^a + \frac{v^{\rm eff}(\rho)}{16\,D_{\rm r}} \, \nabla^2\,\Big(v^{\rm eff}(\rho)\,\bm{w}^a \Big) \\
			&- \sum_{b,c} \frac{ g'_{ab}\,g'_{ac}}{2\,D_{\rm r}} \, \bm{w}^a \, (\bm{w}^b \cdot \bm{w}^c) \\
			&- \frac{z}{16\,D_{\rm r}} \, \nabla\rho \cdot \big[\nabla \big(v^{\rm eff}(\rho)\,\bm{w}^a\big) - \nabla^* \big(v^{\rm eff}(\rho)\,\bm{w}^{a*} \big) \big]\\
			&+ \sum_b \frac{g'_{ab}}{8 \,D_{\rm r}} \Big[ \bm{w}^b \cdot \nabla \big(v^{\rm eff}(\rho)\, \bm{w}^a\big) \\
			& \qquad + \bm{w}^{b*} \cdot \nabla \big(v^{\rm eff}(\rho)\, \bm{w}^{a*}\big) -2 \, \Big\{ v^{\rm eff}(\rho)\,\bm{w}^a \cdot \nabla \bm{w}^b \\
			&\qquad   +\bm{w}^b \, \nabla \cdot \big(v^{\rm eff}(\rho)\,\bm{w}^a\big) - v^{\rm eff}(\rho)\, \bm{w}^{a*} \cdot \nabla \bm{w}^{b*} \\
			&\qquad - \bm{w}^{b*}\,  \nabla \cdot \big( v^{\rm eff}(\rho)\, \bm{w}^{a*}\big) \Big \}  \Big] .
		\end{split}
\end{equation} }

As explained in \cite{kreienkamp_klapp_2022_clustering_flocking_chiral_active_particles_non-reciprocal_couplings}, the density flux $\bm{j}_a$, given in equation \eqref{eq:continuity_flux}, reflects that the motion of particles belonging to species $a$ in space is a result of their self-propulsion in the direction $\bm{w}^a$. The self-propulsion velocity is not constant but particles are slowed down in crowded situations due to the density-dependent velocity $v^{\rm eff}(\rho)={\rm Pe}- z\,\rho$ with $\rho=\sum_b\rho^b$. Additionally, the flux comprises translational diffusion. 
The evolution of the polarization density $\bm{w}^a$, as described by equation \eqref{eq:continuum_eq_polarization}, stems from various contributing factors. These include the tendency of particles to swim (with increasing speed) towards low-density regions (first term on right-hand side), the decay of the polarization due to rotational diffusion (second term), and the orientational coupling of particles among all species (third term). The remaining (diffusional and non-linear) terms contribute to the smoothing out of low- and high-polarization regions.

In Eq.~\eqref{eq:continuum_eq_polarization}, we have introduced $\bm{w}^*=(w_y, -w_x)^{\rm T}$ and $\nabla^*=(\partial_y, -\partial_x)^{\rm T}$. The equations are non-dimensionalized with a characteristic time scale $\tau$ and a characteristic length scale $\ell$. The particle and polarization densities of species $a$ are scaled with the average particle density $\rho_0^a$. The remaining five dimensionless control parameters are the P\'eclet number ${\rm Pe}=v_0\,\tau/\ell$, $z = \zeta\,\rho^a_0\,\tau/\ell$ measuring the particle velocity-reduction due to the environment, the translational diffusion coefficient $D_{\rm t}=\xi\,\tau/\ell^2$, the rotational diffusion coefficient $D_{\rm r}=\eta\,\tau$, and $g'_{ab} = k_{ab}\,\mu_{\theta}\,R_{\theta}^2\,\rho_0^b\,\tau/2$ as relative orientational coupling parameter. Thereby, $g'_{ab}>0$ leads to an alignment and $g'_{ab}<0$ to an anti-alignment of particles. The five control parameters are summarized in Table \ref{tab:control_parameters}.

\begin{table}
	\begin{tabular}{lll}
		\hline
		parameter & definition & description\\
		\hline
		${\rm Pe}$ & $v_0\,\tau/\ell$ & P\'eclet number\\
		$z$ & $\zeta\,\rho^a_0\,\tau/\ell$ & particle velocity-reduction\\
		$D_{\rm t}$ & $\xi\,\tau/\ell^2$ & translational diffusion\\
		$D_{\rm r}$ & $\eta\,\tau$ & rotational diffusion\\
		$g'_{ab}$ & $k_{ab}\,\mu_{\theta}\,R_{\theta}^2\,\rho_0^b\,\tau/2$ & orient. coupling strength\\
		\hline
	\end{tabular}
	\caption{\label{tab:control_parameters}The five control parameters in the non-dimensionalized continuum description \eqref{eq:continuum_eq_density} -- \eqref{eq:continuum_eq_polarization}. The characteristic time and length scales are $\tau$ and $\ell$. The average density is $\rho_0 = \sum_a \rho_0^a$, where $\rho_0^a$ denotes the density of species $a$.}
\end{table}

\subsection{\label{sapp:parameter_choice_continuum}Parameter choice with respect to particle-based model}
In our continuum model, we can adopt many parameters directly from the considered particle simulation parameters. These comprise the P\'eclet number, ${\rm Pe}=40$, and the rotational diffusion constant, $D_r = \eta\,\tau = {3 \cdot 2^{-1/3}}$. The area fraction in particle simulations, $\Phi = 0.4$, transforms into the number density $\rho_0 = 2\, \rho_0^a = 4/\pi\,\Phi$, where $\rho^a_0 = 2/\pi\,\Phi$.  The orientational couplings in continuum simulations ($g'_{ab}$) are related to those in the particle simulations ($g_{ab}$) via $g_{ab}'=0.51\,g_{ab}$, given $R_{\theta}=2\,\ell$. In this study, we focus on systems with fixed weak intraspecies coupling strengths, $g_{AA}=g_{BB}=3$. The interspecies coupling strengths $g_{AB}$ and $g_{BA}$ are chosen independently.   

Nevertheless, there are two parameters that require special attention: the velocity reduction parameter, $\zeta$, and the translational diffusion constant, $D_{\rm t}$.

The parameter $\zeta$ [that arises in Eq.~\eqref{eq:effective_velocity}] is determined by considering the pair distribution function $G_{ab}(r,\theta)$ [Eq.~\eqref{eq:pair_distribution_function}] and evaluating the integral \cite{Bialke_2013_microscopic_theory_phase_seperation, speck_2015_dynamical_mean_field_phase_separation, elena_2021_phase_separation_self-propelled_disks}
\begin{equation}
	\zeta = \int_0^{\infty} {\rm d}r \, r \, \left[ -\mu_r\,U'(r)\right] \int_0^{2\,\pi} {\rm d}\theta \, \cos(\theta) \, G_{ab}(r,\theta) ,
\end{equation}
where $\mu_r$ is the spatial mobility and $U'(r)$ represents the derivative of the WCA potential \eqref{eq:WCA_potential} with respect to the interparticle distance, $r$. To isolate the effect of steric repulsion and to exclude the effect of (anti-)alignment, we simulate a single-species ABP system without orientational couplings, while keeping the other parameters the same. We obtain the non-dimensionalized velocity reduction parameter $z = \zeta \, \rho^a_0 \, \tau/\ell = 57.63 \, \rho^a_0 \, \tau/\ell = 0.37 \, {\rm Pe}/\rho_0^{\rm con}$ with $\rho_0^{\rm con} = 1$. This choice places the system well within the MIPS instability region for a wider range of alignment strengths \cite{kreienkamp_klapp_2022_clustering_flocking_chiral_active_particles_non-reciprocal_couplings}.

The second parameter requiring special attention is the translational diffusion coefficient, $D_{\rm t}$, within the continuum model. Notably, $D_{\rm t}$ does \textit{not} correspond to the quantity $\xi \, \tau/\ell^2$ (with $\xi=1$ in our particle simulations). In previous studies of repulsive active Brownian particles \cite{Bialke_2013_microscopic_theory_phase_seperation,speck_2015_dynamical_mean_field_phase_separation,elena_2021_phase_separation_self-propelled_disks}, $D_{\rm t}$ has been related to the (constant) long-time diffusion coefficient of the passive particle system, which is generally lower than that of the active system. Yet, other studies have shown that the diffusion constant in the continuum model needs to depend on the local density to achieve quantitative agreement with simulations \cite{stenhammar_cates_2013_continuum_theroy_ABP,cates_tailleur_2013_ABP_run-and-tumble_equivalent}. Additionally, our particle-based simulations reveal that the diffusion constant depends significantly on the strength and non-reciprocity of (anti-)alignment couplings. 

To keep the calculations treatable, we make an ad-hoc choice of $D_{\rm t}=9$ in our continuum description. This choice enables the continuum model to capture qualitatively the various observed behaviors of the system at different orientational coupling strengths. Smaller or larger values of $D_{\rm t}$ lead qualitatively to the same behavior on the coarse-grained level of description. The precise value only affects the location of the transition from the asymmetrical clustering to the disordered state in the $g_{AB}-g_{BA}$-plane.

\subsection{\label{sapp:continuum_simulations}Continuum simulations}
In this study, we use the mean-field continuum model mainly as a starting point for a linear stability analysis, see Appendix \ref{app:linear_stability_analysis}. To test the validity of this analysis, we have also performed numerical simulations of the full continuum Eqs.~\eqref{eq:continuum_eq_density}–\eqref{eq:continuum_eq_polarization} in two-dimensional periodic systems.

\begin{figure*}
	\includegraphics[width=1\linewidth]{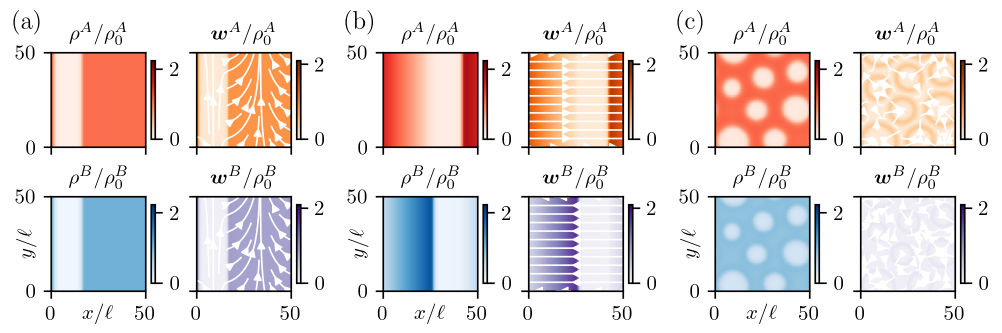}
	\caption{\label{fig:continuum_simulations}Continuum simulations. (a) $g_{AB}=g_{BA}=9$. (b) $g_{AB}=g_{BA}=-9$. (c) $g_{AB}=-g_{BA}=9$. The patterns are stationary. Other parameters are chosen as described in Appendices \ref{sapp:parameter_choice_continuum} and \ref{sapp:continuum_simulations}.}
\end{figure*}

To this end, we employ a pseudo-spectral code combined with an operator splitting technique, allowing us to accurately treat the linear operator using a fourth-order Runge Kutta time integration scheme \cite{canuto_zang_2007_spectral_methods}. The initial state is chosen as a slightly perturbed disordered state with zero polarization $\bm{w}(\bm{r},0)=\bm{0}$ and a constant density $\rho^a(\bm{r},0) = 1$. The two-dimensional simulation box of size $50 \times 50$ is discretized into $125 \times 125$ grid points. Other parameters are specified in Appendix \ref{sapp:parameter_choice_continuum}.

Snapshots of the continuum simulations corresponding to three different interspecies coupling strengths $g_{AB}$, $g_{BA}$ are shown in Fig.~\ref{fig:continuum_simulations}. In case of strong reciprocal alignment with $g_{AB}=g_{BA}=9$, the continuum simulations demonstrate the formation of high-density bands where particles of both species occupy the same space and align in the same direction. This is characteristic for a flocking state [Fig.~\ref{fig:continuum_simulations}(a)]. Conversely, for strong reciprocal anti-alignment, particles of both species accumulate in separate high-density regions. Within these regions, particles of the same species align and form anti-parallel flocks that face the flock of the other species [Fig.~\ref{fig:continuum_simulations}(b)].
In case of non-reciprocal (anti-)alignment couplings of strengths $g_{AB}=-g_{BA}=9$, the continuum simulations evolve towards a pattern of co-existing higher and lower density regions [Fig.~\ref{fig:continuum_simulations}(c)]. Both species $A$ and $B$ accumulate in roughly the same regions, without significant emerging polarization. Locally, the density of species $A$ is larger than of species $B$, such that one may conclude that cluster formation is increased for species $A$ as compared to $B$. Importantly, the continuum simulations of these three scenarios yield \textit{stationary} patterns, in contrast to the dynamical phases observed in particle simulations.\\

\section{\label{app:linear_stability_analysis}Mean-field linear stability analysis}
As a major tool to explore the overall phase behavior, we investigate the linear stability of the disordered, uniform state characterized by a uniform density and zero polarization for both species $a=A,B$, i.e., $(\rho^a, \bm{w}^a)=(1,\bm{0})$. Remember that Eqs.~\eqref{eq:continuum_eq_density}–\eqref{eq:continuum_eq_polarization} are already scaled with the mean density $\rho_0^a$. The linear stability analysis can be done analytically. To this end, we consider perturbations to the disordered state involving all wave vectors $\bm{k}$,
\begin{subequations} \label{eq:perturbation_form}
	\begin{align}
		\rho^{a\prime}(\bm{r},t) &= \int \hat{\rho}^a(k) \, {\rm e}^{i\bm{k}\cdot \bm{r} + \sigma(k)t}\, {\rm d}\bm{k}\\
		\bm{w}^{a\prime}(\bm{r},t) &= \int \bm{\hat{w}}^a(k) \, {\rm e}^{i\bm{k}\cdot \bm{r} + \sigma(k)t}\, {\rm d}\bm{k}.
	\end{align}
\end{subequations}
In this way, perturbations are expressed as plane waves with a (complex) growth rate $\sigma(k)$  and amplitudes $\hat{\rho}^a(k)$ and $\bm{\hat{w}}^a(k)$. Here, $\sigma$ depends only on the wave number $k = \vert \bm{k} \vert$, because we study the stability of the isotropic base state.

\begin{table*}
	\begin{tabular}{| p{2.2cm} | p{6.3cm} | p{6cm} |}
		\hline
		non-eq.\ state & eigenvalues $\sigma_i$ & eigenvector $\bm{v}$ of largest real eigenvalue\\
		\hline
		\hline
		disorder & ${\rm Re}(\sigma_i(k))\leq 0$ for all $k$ and $i=0,...,6$ & --\\
		\hline
		flocking & ${\rm Re}(\sigma_i(k=0))> 0$ for any $i$ & largest entries of eigenvector in $\hat{\bm{w}}^A + \hat{\bm{w}}^B$\\
		\hline
		anti-flocking & ${\rm Re}(\sigma_i(k=0))> 0$ for any $i$ & largest entries of eigenvector in $\hat{\bm{w}}^A - \hat{\bm{w}}^B$\\
		\hline
		sym.\ clustering & ${\rm Re}(\sigma_i(k=0))\leq 0$ for all $i$ and global maximum ${\rm Re}(\sigma_i(k_{\rm max}))$ at $k_{\rm max}>0$ for any $i$ & $\alpha \approx 0$\\
		\hline
		de-mixing & ${\rm Re}(\sigma_i(k=0))\leq 0$ for all $i$ and global maximum ${\rm Re}(\sigma_i(k_{\rm max}))$ at $k_{\rm max}>0$ for any $i$ & $\alpha \approx \pm\pi/2$\\
		\hline
		asym.\ cl.\ $A$ & ${\rm Re}(\sigma_i(k=0))\leq 0$ for all $i$ and global maximum ${\rm Re}(\sigma_i(k_{\rm max}))$ at $k_{\rm max}>0$ for any $i$ & $0<\alpha < \pi/2$\\
		\hline
		asym.\ cl.\ $B$ & ${\rm Re}(\sigma_i(k=0))\leq 0$ for all $i$ and global maximum ${\rm Re}(\sigma_i(k_{\rm max}))$ at $k_{\rm max}>0$ for any $i$ & $-\pi/2<\alpha < 0$\\
		\hline
	\end{tabular}
	\caption{\label{tab:phases_eigenvalues_eigenvectors_characterization}Characterization of non-equilibrium states in the repulsive binary mixture with non-reciprocal orientational alignment couplings. The six eigenvalues and the eigenvector corresponding to the largest eigenvalue determine the non-equilibrium state. The angle $\alpha=\arccos(\bm{v}_{\rho} \cdot \bm{x}_{\rm con})$ with $\bm{v}_{\rho}=(\hat{\rho}^A+\hat{\rho}^B, \, \hat{\rho}^A-\hat{\rho}^B)^{\rm T}$ and $\bm{x}_{\rm con}=(1,0)^{\rm T}$ indicates the type of phase transition.}
\end{table*}  

As we consider a binary mixture of species and are interested in the collective dynamics of the species with respect to each other, we look at perturbations in the combined field quantities $\rho^A+\rho^B$ (total density), $\rho^A-\rho^B$ (density difference), $\bm{w}^A+\bm{w}^B$ (total polarization), and $\bm{w}^A-\bm{w}^B$ (polarization difference). 

To this end, we insert the ansatz $\rho^a(\bm{r},t) = 1 + \rho^{a\prime}(\bm{r},t)$, $\bm{w}^a(\bm{r},t) = \bm{w}^{a\prime}(\bm{r},t)$ into the time evolution equations for $\rho^A+\rho^B$, $\rho^A-\rho^B$, $\bm{w}^A+\bm{w}^B$, and $\bm{w}^A-\bm{w}^B$, which are readily obtained from Eqs.~\eqref{eq:continuum_eq_density} - \eqref{eq:continuum_eq_polarization}. We then assume $\rho^{a\prime}$ and $\bm{w}^{a\prime}$ to be small. Linearization leads to a system of equations that is decoupled with respect to wave number $k$. For each $k$, we arrive at an eigenvalue equation
\begin{equation}
	\label{eq:eigenvalue_equation}
	\sigma(k) \, \bm{v}(k) = \bm{\mathcal{M}}(k) \cdot \bm{v}(k).
\end{equation}
Here, the 6-component eigenvector $\bm{v}(k) = (\hat{\rho}^A+\hat{\rho}^B, \, \hat{\rho}^A-\hat{\rho}^B,\, \hat{w}_x^A + \hat{w}_x^B,\, \hat{w}_y^A + \hat{w}_y^B,\,  \hat{w}_x^A - \hat{w}_x^B,\, \hat{w}_y^A - \hat{w}_y^B)^{\rm T}$ contains the possible perturbations of the particle densities and the two components of the polarization densities. The $6 \times 6$ matrix $\bm{\mathcal{M}}(k)$ is given by
\begin{widetext}
	\begin{equation}
		\label{eq:linear_stability_matrix}
		\bm{\mathcal{M}}(k) = \begin{pmatrix}
			-D_{\rm t}\,k^2 & 0 & -i\,V\,k_x & -i\,V\,k_y & 0 & 0\\
			0 & -D_{\rm t}\,k^2 & 0 & 0 & -i\,V\,k_x & -i\,V\,k_y\\
			-\tfrac{i}{2}(V-2\,z)\,k_x & 0 & C_{++}-D_{w} & 0 & C_{+-} & 0 \\
			-\tfrac{i}{2}(V-2\,z)\,k_y & 0 & 0& C_{++}-D_{w} & 0 & C_{+-}  \\
			0 & -\tfrac{i}{2}V\,k_x & C_{-+} & 0 & C_{--} -D_{w} & 0 \\
			0 & -\tfrac{i}{2}V\,k_y & 0 & C_{-+} & 0 & C_{--} -D_{w} \\
		\end{pmatrix},
	\end{equation}
\end{widetext}
where $V = {\rm Pe} - 2\,z$ and $D_w =  (V^2/(16\,D_r) + \mathcal{D}_{\rm t}) \, k^2 + D_r$. 
The orientational couplings are given by
\begin{equation}
	C_{++} = \tfrac{1}{2}(g'_{AA}+g'_{AB}+g'_{BA}+g'_{BB}),
\end{equation}
\begin{equation}
	C_{+-} = \tfrac{1}{2}(g'_{AA}-g'_{AB}+g'_{BA}-g'_{BB}),
\end{equation}
\begin{equation}
	C_{-+} = \tfrac{1}{2}(g'_{AA}+g'_{AB}-g'_{BA}-g'_{BB}),
\end{equation}
\begin{equation}
	C_{--} = \tfrac{1}{2}(g'_{AA}-g'_{AB}-g'_{BA}+g'_{BB}).
\end{equation}
From Eq.~\eqref{eq:eigenvalue_equation}, we can derive analytical expressions for the (complex) growth rates $\sigma(k)$, which play the roles of eigenvalues. We mainly focus on the real part of the eigenvalues, ${\rm Re}(\sigma)$, which determines the actual growth or decay of the perturbations in time. The imaginary parts are related to oscillatory behavior, which is essentially absent at the parameters studied in this work. The disordered state becomes linearly unstable if ${\rm Re}(\sigma(k)) > 0$ for any $k$. We monitor all six functions ${\rm Re}(\sigma(k))$ and analyze the largest value and corresponding eigenvector, which determine the type of emerging dynamics at short times \cite{kreienkamp_klapp_2022_clustering_flocking_chiral_active_particles_non-reciprocal_couplings}.

At $k=0$, the eigenvalues read
\begin{subequations} \label{eq:eigenvalues_k-0}
	\begin{align}
		\sigma_{1/2}(k=0) &= 0\\
		\sigma_{3/4/5/6}(k=0) &= \frac{g'_{AA}+g'_{BB}}{2} - D_{\rm r}  \\
		& \quad \pm \sqrt{ g'_{AB}\,g'_{BA} + \frac{(g'_{AA}-g'_{BB})^2}{4}} .\notag
	\end{align}
\end{subequations}
The first two growth rates (\ref{eq:eigenvalues_k-0}) vanish due the
conservation of the particle density. The other four growth rates are related to polarization dynamics. They can become positive for strong enough alignment and even imaginary for antagonistic interspecies couplings (i.e., $g'_{AB}\,g'_{BA}<0$).

\subsection{Characterization of emerging states}
The emerging non-equilibrium states can be characterized in terms of eigenvalues and the eigenvector corresponding to the largest eigenvalue.

\begin{figure*}
	\includegraphics[width=\linewidth]{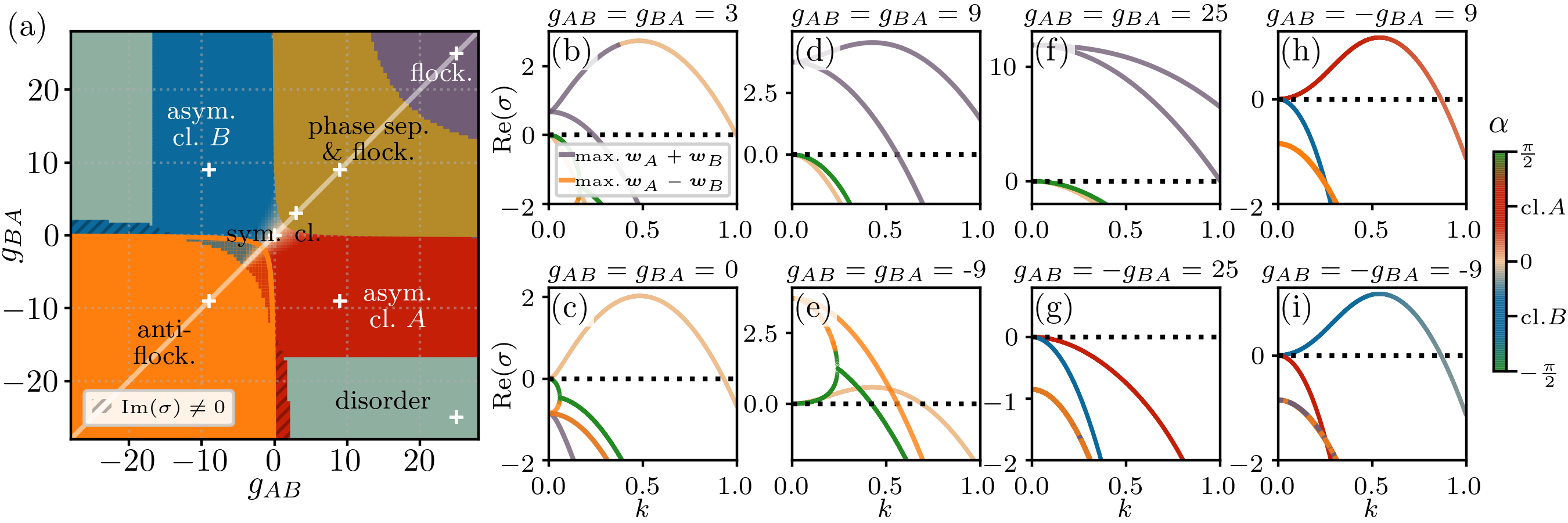}
	\caption{\label{fig:stability_diagram_with_growth_rates}(a) Phase diagram from mean-field stability analysis. (b-i) Growth rates for various parameter combinations. The phases are determined from linear stability analyses of the disordered base state of the continuum Eqs.~\eqref{eq:continuum_eq_density}-\eqref{eq:continuum_eq_polarization}. The white crosses in (a) indicate the parameter combinations whose growth rates are plotted in (b-i). The other parameters are set to $g_{AA}=g_{BB}=3$, ${\rm Pe}=40$, $z=57.63\,\rho_0^a\,\tau/\ell$, $D_{\rm t}=9$, $D_{\rm r} = 3\cdot 2^{-1/3}$, and $\rho_0^a=4/(5\,\pi)$. See also \cite{kreienkamp_klapp_PRL}.}
\end{figure*}

As stated before, the real parts of the (six) eigenvalues, ${\rm Re}(\sigma_i)$ of $\bm{\mathcal{M}}(k)$ [see Eq.~\eqref{eq:linear_stability_matrix}], indicate whether the disordered (base) state is stable or not. As soon as any eigenvalue has a positive real part at any wave number $k$, the system features instabilities. We characterize this as introduced in our earlier work \cite{kreienkamp_klapp_2022_clustering_flocking_chiral_active_particles_non-reciprocal_couplings}. 

If real parts of the eigenvalues become positive at \textit{zero} wave number ($k=0$), we can deduce that the corresponding instabilities concern the polarization dynamics, that is, the emergence of \textit{\mbox{(anti-)}flocking}. The reason is that the growth rate ${\rm Re}(\sigma)$ at $k=0$ determines the growth or decay of spatially integrated fields. The particle density is a conserved quantity, such that the density-associated growth rates must vanish at $k=0$. Hence, the eigenvector $\bm{v}(k=0)$ corresponding to ${\rm Re}(\sigma(k=0)) > 0$ indicates the type of flocking. If the largest entry of the eigenvector occurs in $\bm{\hat{w}}^A + \bm{\hat{w}}^B$ ($\bm{\hat{w}}^A - \bm{\hat{w}}^B$), the linear stability analysis predicts flocking (antiflocking).

Instabilities at finite wavenumbers ($k>0$) pertain to the density dynamics. The eigenvector corresponding to the largest real eigenvalue indicates the type of phase separation. To this end, we consider only the two density-related entries of the (normalized) eigenvector $\bm{v}(k)$, that is, $\bm{v}_{\rho}=(\hat{\rho}^A+\hat{\rho}^B, \, \hat{\rho}^A-\hat{\rho}^B)^{\rm T}$, at small $k>0$. In case of \textit{symmetric clustering}, $\bm{v}_{\rho}=\bm{x}_{\rm con}=(1,0)^{\rm T}$. The angle $\alpha=\arccos(\bm{v}_{\rho} \cdot \bm{x}_{\rm con})$ between $\bm{v}_{\rho}$ and $\bm{x}_{\rm con}$ is approximately $0$. 
In case of \textit{demixing}, $\bm{v}_{\rho}$ is close to $(0,1)^{\rm T}$ with $\alpha\approx \pm \pi/2$. \textit{Asymmetrical clustering} is defined by emerging clusters consisting of mainly one of the two species. For asymmetrical clusters of species $A\ (B)$, the angle is $0<\alpha < \pi/2\ (-\pi/2<\alpha < 0)$.  

Our criteria to characterize the non-equilibrium states are summarized in Table \ref{tab:phases_eigenvalues_eigenvectors_characterization}. Furthermore, Fig.~\ref{fig:stability_diagram_with_growth_rates} shows exemplary real growth rates with indicated largest entries of eigenvectors.

Note that (anti-)flocking and (a)symmetric clustering can occur independent of each other, or in combination. Pure (anti-)flocking is characterized by a global maximum of the growth rate at $k=0$. On the other hand, a combination of (anti-)flocking and (a)symmetric clustering features a positive growth rate at $k=0$, while the maximal growth rate occurs at a finite $k>0$ [see full phase diagram in Fig.~\ref{fig:stability_diagram_with_growth_rates}(a)]. In Fig.~\ref{fig:stability_diagram_with_snapshots}, the (anti-)flocking regions include the $k>0$-instabilities of (a)symmetric clustering.

In our system with relatively weak intraspecies alignment couplings ($g_{AA}=g_{BB}=3$), the eigenvalues are real for the vast majority of intraspecies coupling strengths. Eigenvalues with positive real part and non-zero imaginary part, would indicate oscillatory instabilities. Such behavior is only seen at much larger intraspecies coupling \cite{kreienkamp_klapp_PRL}.

\section{\label{app:asymmetric_clustering_explanation}Asymmetric clustering behavior: microscopic origin and implications on cluster size}
\begin{figure}
	\includegraphics[width=\linewidth]{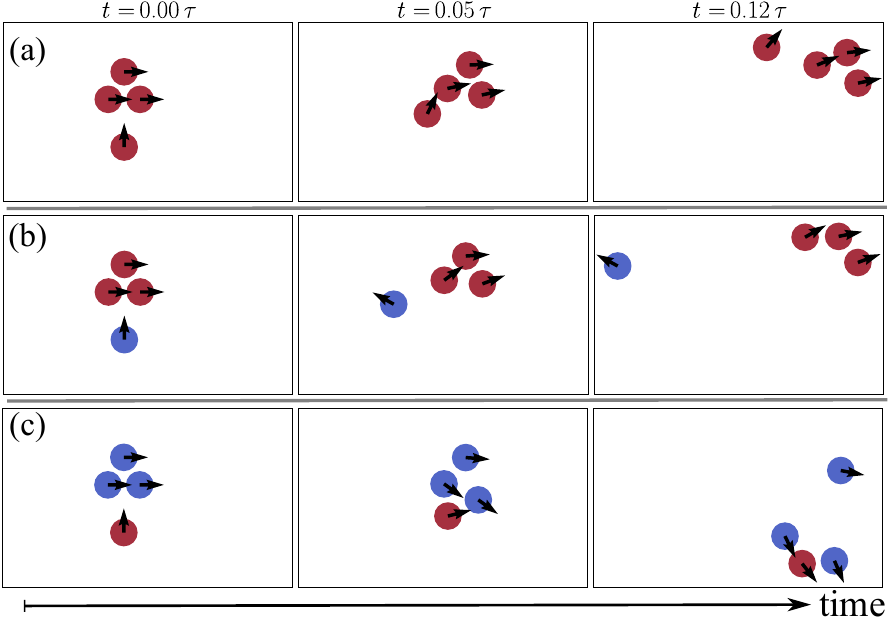}
	\caption{\label{fig:sketch_asymmetric_clustering_N-4}Evolution of small ``clusters'' upon approaching particles of the same or different species. Here, with $g_{AA}=g_{BB}=3$ and $g_{AB}=-g_{BA}=9$, this eventually leads to asymmetric clustering of species $A$. The numerical simulations are shown for the noiseless limit. Particles of species $A$ $(B)$ are colored in red (blue). (a,b) Coherently moving clusters of species $A$ survive the approach of $A$- and $B$-particles. (c) $B$-clusters are destabilized by approaching $A$-particles.}
\end{figure}

To shed light on the microscopic origin of the asymmetric clustering behavior caused by non-reciprocal orientational couplings, we consider the exemplary situation of $g=3$ and $\delta=9$. In this case, particles of species $A$ want to align with other $A$- and $B$-particles. On the other hand, particles of species $B$ want to align only with other $B$-particles and orient opposite to $A$-particles. In Fig.~\ref{fig:sketch_asymmetric_clustering_N-4} we illustrate how, in this case, $A$-clustering is stabilized while $B$-clustering is de-stabilized. The main argument has been outlined in \cite{kreienkamp_klapp_PRL} and is briefly summarized here. 
Generally, when particles align and start to move coherently, clustering of these particles is enhanced \cite{martin_gomez_pagonabarraga_2018_collective_motion_ABP_alignment_volume_exclusion,elena_2018_velocity_alignment_MIPS}. Consider now one of such small, coherently moving ``clusters'' consisting of either three $A$- or $B$-particles upon an approaching fourth particle of either the same or different species. When the $A$($B$)-cluster is approached by another $A$($B$)-particle [case Fig.~\ref{fig:sketch_asymmetric_clustering_N-4}(a)], this particle either joins the already coherently moving cluster or, at least, does not significantly disturb its motion, depending on the initial configuration. When an $A$-cluster is approached by a $B$-particle [case Fig.~\ref{fig:sketch_asymmetric_clustering_N-4}(b)], the latter re-orients into the opposite direction of the coherent $A$-motion ($g_{BA}<0$), moving away from the $A$-cluster. The $A$-cluster is not significantly disturbed by the quickly departing $B$-particle. On the other hand, when a $B$-cluster is approached by an $A$-particle [case Fig.~\ref{fig:sketch_asymmetric_clustering_N-4}(c)], the $A$-particle tends to orient along the cluster's direction ($g_{AB}>0$). At the same time, only $B$-particles close to the approaching particle (but not all $B$-particles) re-orient into the opposite direction of the $A$-particle ($g_{BA}<0$). Since intraspecies couplings are relatively weak ($g<\delta$), this eventually leads to diverging trajectories of the originally clustered $B$-particles and the destruction of the original cluster.
Hence, small clusters of species $A$ are less susceptible to disturbances, while species-$B$ clusters easily get destructed by other $A$-particles.

These microscopic considerations also help us understand how the largest cluster size is affected when deviating from the fully anti-symmetric case ($g_{AB}=-g_{BA}$). As shown in Fig.~\ref{fig:largest_cluster_size_asymmetry} and snapshot Fig.~\ref{fig:stability_diagram_with_snapshots}(h), cluster formation of $A$ is even more pronounced for $g_{AB}=6, g_{BA}=-9$ than for $g_{AB}=-g_{BA}=-9$. The reason behind is that, for $g_{AB}=6$, $A$ is less ``distracted'' by $B$ than for $g_{AB}=9$. At the same time, $B$ still anti-aligns as strongly as before. On the other hand, asymmetric $A$-clustering is less pronounced for $g_{AB}=9, g_{BA}=-6$, where $A$ strongly aligns with $B$-particles and easily gets distracted by them, while $B$ anti-aligns less strongly and does not move away as quickly. 

\section{\label{app:analysis_methods_details_clustering}Clustering and phase separation in particle simulations}
To determine whether the system is in a phase-separated state, we employ the position-resolved local area fraction, whose distribution exhibits a double-peak structure when dense and dilute regions coexist \cite{blaschke_2016_phase_separation,Liao_2018_circle_swimmers_monolayer,liao_2020_dynamical_self-assembly_dipolar_active_Brownian_particles}.
The position-resolved local area fraction is determined in two steps. First, we use Voronoi cells to assign a local area to every particle in the system. The particle-resolved local area fraction of particle $i$ is given by
\begin{equation}
	\Phi_i = \frac{\pi\,\sigma^2}{4\,A_i}, 
\end{equation} 
where $A_i$ is the area of the Voronoi cell associated with particle $i$. In a second step, the the particle-resolved local area fractions are mapped onto the position-resolved local area fraction $\Phi(x,y)$ using a grid of mesh size $\Delta L = L/{\rm floor}(L) \approx 1\,\sigma$ \footnote{The floor function ${\rm floor}(x)$ yields the largest integer less than or equal to any real number $x$.}. At grid points $(x,y)$ within the Voronoi cell $A_i$, the position-resolved local area fraction is assigned to the value of the particle-resolved local area fraction, i.e., $\Phi(x,y)\vert_{(x,y)\in A_i} = \Phi_i$. Also within a non-equilibrium steady state, the local area fraction fluctuates over time. We thus calculate the time average $\overline{\Phi}(x,y) = \langle \Phi(x,y) \rangle_t$ between $98$ and $100\,\tau$ after initialization.

To characterize the clusters within the system quantitatively, we determine the largest cluster size. Clusters are identified using a distance criterion, where particles $i$ and $j$ are considered to be in contact if $r_{ij}<r_{\rm c}$ \cite{buttinoni_2013_dynamical_clustering, Bialke_2013_microscopic_theory_phase_seperation,Liao_2018_circle_swimmers_monolayer}. A cluster is defined as the set of particles that are in contact with each other. The cluster size $n$ represents the number of particles within the cluster. We distinguish between clusters made of particles of any species and cluster made of particles of one species $A$ or $B$.
Assuming that, at a given time $t$, there are $N_{\rm c}(t)$ clusters with respective sizes $n_i(t)$, $1\leq i \leq N_{\rm c}(t)$, the instantaneous largest cluster size $n_{\rm lcl}(t)$ is defined as the largest number among $n_i(t)$. The time-averaged largest cluster size is $\langle n_{\rm lcl} \rangle_t$, whereby the time average is taken between $70$ and $120\,\tau$ after initialization. We define the ratio of all particles in the largest cluster as $\mathcal{N}_{\rm lcl} = \langle n_{\rm lcl} \rangle_t/N$ and the ratio of particles of species $a$ in a pure species-$a$ cluster as $\mathcal{N}_{{\rm lcl},a} = \langle n_{{\rm lcl},a} \rangle_t/N_a$.


%

\end{document}